\documentclass{tlp}

\usepackage{aopmath}
\usepackage{graphics}
\usepackage{color}
\usepackage{graphicx}
\usepackage{natbib}

\newtheorem{definition}{Definition} 
\newtheorem{example}{Example}

\def\<{\mbox{$\langle$}}
\def\>{\mbox{$\rangle$}}
\newcommand{\st}{\;|\,}

\newcommand{\U}{\mathcal{U}}
\newcommand{\V}{\mathcal{V}}
\newcommand{\A}{\mathcal{A}}
\newcommand{\E}{\mathcal{E}}
\newcommand{\F}{\mathcal{F}}
\newcommand{\W}{\mathcal{W}}

\newcommand{\act}{\mathit{ua}}
\newcommand{\upd}{\mathit{up}}
\newcommand{\ua}{\mathit{ua}}
\newcommand{\aic}{\mathit{AIC}}
\newcommand{\prop}{\mathit{prop}}
\newcommand{\nup}{\mathit{nup}}

\newcommand{\nt}{\mathit{not\,}}
\newcommand{\nef}{\mathit{ne}}

\newcommand{\lit}{\mathit{lit}}

\newcommand{\I}{\mathcal{I}}
\newcommand{\J}{\mathcal{J}}
\newcommand{\R}{\mathcal{R}}
\newcommand{\bR}{\mathbf{R}}
\newcommand{\bFR}{\mathbf{FR}}
\newcommand{\bJR}{\mathbf{JR}}
\newcommand{\bWR}{\mathbf{WR}}
\newcommand{\bFWR}{\mathbf{FWR}}
\newcommand{\bJWR}{\mathbf{JWR}}
\newcommand{\bRev}{\mathbf{Rev}}
\newcommand{\bFRev}{\mathbf{FRev}}
\newcommand{\bJRev}{\mathbf{JRev}}
\newcommand{\bWRev}{\mathbf{WRev}}
\newcommand{\bFWRev}{\mathbf{FWRev}}
\newcommand{\bJWRev}{\mathbf{JWRev}}

\newcommand{\At}{\mathit{At}}
\newcommand{\bd}{\mathit{body}}
\newcommand{\hd}{\mathit{head}}

\newcommand{\rin}{\mbox{\bf{in}}}
\newcommand{\rout}{\mbox{\bf{out}}}
\newcommand{\vinc}{\rotatebox[origin=c]{-90}{$\subseteq$}}
\newcommand{\veq}{\rotatebox{90}{$=$}}

\begin{document}
\bibliographystyle{acmtrans}

\long\def\comment#1{}

\title{Active Integrity Constraints and Revision Programming}

\author[L. Caroprese and M. Truszczy{\'n}ski]
{Luciano Caroprese$^1$ and
Miros{\l}aw Truszczy{\'n}ski$^2$ \\
$^1$Universit\`a della Calabria\\
87030 Rende, Italy\\
E mail: caroprese@deis.unical.it and\\
$^2$Department of Computer Science, University of Kentucky\\
Lexington, KY~40506, USA\\
E mail: mirek@cs.uky.edu}

\pagerange{\pageref{firstpage}--\pageref{lastpage}}
\setcounter{page}{1}
\pubyear{2009}

\submitted{25 April 2009}
\revised{18 November 2009, 17 May 2010}
\accepted{3 August 2010}
 
\maketitle

\label{firstpage}

\begin{abstract}
We study \emph{active integrity constraints} and \emph{revision programming},
two formalisms designed to describe integrity constraints on databases and 
to specify policies on \emph{preferred} ways to enforce them. Unlike other
more commonly accepted approaches, these two formalisms attempt to provide 
a declarative solution to the problem. However, the original semantics of 
\emph{founded repairs} for active integrity constraints and \emph{justified
revisions} for revision programs differ. Our main goal is to establish a 
\emph{comprehensive framework} of semantics for active integrity constraints, 
to find a parallel framework for revision programs, and to relate the two.
By doing so, we demonstrate that the two formalisms proposed independently 
of each other and based on different intuitions when viewed within a broader 
semantic framework turn out to be notational variants of each other. That 
lends support to the adequacy of the semantics we develop for each of the 
formalisms as the foundation for a declarative approach to the problem of 
database update and repair. In the paper we also study computational
properties of the semantics we consider and establish results concerned
with the concept of the minimality of change and the invariance under
the shifting transformation.

\vspace{2mm}
\noindent
\emph{To appear in Theory and Practice of Logic Programming (TPLP)}
\end{abstract}

 
\section{Introduction}
Integrity constraints are conditions on databases. If a database 
violates integrity constraints, it needs to be \emph{repaired} --- 
updated so that the integrity constraints hold again. Often there are 
several ways to enforce integrity constraints. The paper
is concerned with the problem to \emph{specify} policies for preferred ways 
to repair databases in a \emph{declarative} way as part of the description of 
integrity constraints.

A database can be viewed as a finite set of ground atoms in the
language of first-order logic determined by the database schema and an
infinite countable set of constants. An integrity constraint can be 
modeled by a formula in this language. A database \emph{satisfies} 
an integrity constraint if it is its \emph{Herbrand} model. Since 
databases and sets of integrity constraints are \emph{finite}, without
loss of generality, we will limit our attention to the case when
databases are subsets of some finite set $\At$ of \emph{propositional}
atoms, and integrity constraints are clauses in the propositional 
language generated by $\At$. The notions we propose and 
the results we obtain in that restricted setting lift to the first-order 
one (including aggregate operations and built-in predicates) via the 
standard concept of \emph{grounding}. We do not discuss this matter here
in more detail, as our main objective is to develop a semantic framework 
for declarative specifications of repair policies rather than to study
practical issues of possible implementations.

To illustrate the problem of database repair with respect to integrity
constraints, let us consider the database $\I=\{a,b\}$ and the integrity
constraint $\neg a \vee \neg b$. Clearly, $\I$ does not satisfy $\neg a
\vee \neg b$ and needs to be ``repaired'' --- replaced by a database that
satisfies the constraint. Assuming $\At=\{a,b,c,d\}$, the databases 
$\emptyset$, $\{a\}$, $\{b\}$, $\{a,c\}$ are examples of databases that 
could be considered as replacements for $\I$. Since the class of
replacements of $\I$ is quite large, the question arises whether there
is a principled way to narrow it down. One of the most intuitive and 
commonly accepted postulates is that the change between the initial 
database $\I$ and the revised database $\R$, given by $\I\div\R$, be
minimal (an example of an early work exploiting that idea is the paper
by Winslett \citeyear{win90}; for a more detailed discussion of
the role of minimality in studies of database updates we refer to the 
paper by Chomicki \citeyear{Chomicki07}). In our case, the minimality of 
change narrows down the class of possible revisions to $\{a\}$ and 
$\{b\}$. 

In some cases, the minimality of change is not specific enough and may 
leave too many candidate revisions. The problem can be addressed by 
formalisms that allow the database designer to formulate integrity 
constraints and, in addition, to state preferred ways for enforcing them.
In this paper, which represents an extended version of two conference 
papers \citep{CaroTrusJELIA,CaroTrusICLP}, we study two such formalisms: 
active integrity constraints introduced by Caroprese, Greco, Sirangelo 
and Zumpano \citeyear{CaroGZ}, and revision programming introduced by
Marek and Truszczy\'nski \citeyear{mt94c}.

Active integrity constraints and revision programs are
languages for specifying integrity constraints. However, unlike in the
standard case, when integrity constraints are just first-order formulas
that make no distinctions among its models, both sets of active integrity 
constraints and revision programs are meant to represent \emph{policies}
for preferring some models over others. In other words, they give database 
system
designers means to express policies for narrowing down the space of models
that need to be considered when repairing inconsistencies or when 
querying an inconsistent database. In a sense, the two formalisms arise 
from the need to provide declarative counterparts to procedural attempts 
to accomplish the same objective \citep{WidomC96c,jagadish99managing}. 
An in-depth understanding of the semantics
and, in general, properties of these two formalisms is then essential. 
Developing that understanding is the main goal of our paper. 

To recall, active integrity constraints \emph{explicitly} encode both 
integrity constraints and preferred basic actions to repair them, in 
the case
when the constraints are violated. To specify a precise meaning of sets
of active integrity Caroprese et al. \citeyear{CaroGZ} 
proposed the semantics of \emph{founded repairs}.
Founded repairs are change-minimal and satisfy a certain groundedness
condition.

Revision programs consist of \emph{revision rules}. Each revision rule 
represents an integrity constraint, and \emph{implicitly} encodes 
preferred ways to enforce it by means of a certain syntactic convention.
Following intuitions from logic programming, Marek and Truszczy\'{n}ski 
\citeyear{mt94c} proposed two semantics for revision programs: the 
semantics of \emph{justified} revisions and the semantics of \emph{supported} 
revisions. Each semantics reflects some form of preferences on ways to 
repair a database given a revision program. 

The original semantics of active integrity constraints and revision
programming seemingly cannot be related in any direct way. They have 
different computational properties. For instance, the problem of the 
existence of a founded repair for a set of normal active integrity constraints 
is $\Sigma_P^2$-complete, while the same problem for justified revisions 
of normal revision programs is NP-complete. Furthermore, 
while the semantics
for revision programming do not have the minimality of change property,
founded repairs with respect to active integrity constraints do.

In this paper, we demonstrate that despite the differences in the 
syntax, and the lack of a simple correspondence between justified 
revisions and founded repairs, the formalisms of revision programs and 
active integrity constraints are closely related. There are two keys
to the relationship. First, we need a certain syntactic restriction on 
revision programs. Specifically, we introduce the class of \emph{
proper} revision programs and show that restricting to proper programs
does not affect the expressive power. 

Second, we need to broaden the families of the semantics for each
formalism so that the two sides could be aligned. To this end
for active integrity constraints we introduce new semantics
by dropping the minimality of change condition, which results in the
semantics of \emph{weak repairs} and \emph{founded weak repairs}. We
also adapt to the case of active integrity constraints the semantics of
justified revisions (justified weak revisions), which leads us to the
semantics of \emph{justified weak repairs} and \emph{justified repairs}.
For revision programs, we modify the semantics of revisions and justified revisions 
by imposing on them the minimality
condition. Moreover we introduce the semantics of
\emph{founded revisions} (\emph{founded weak revisions}) that corresponds 
to the semantics of founded repairs (founded weak repairs).
We show that under a simple bijection between proper revision programs
and active integrity constraints, founded (weak) revisions correspond to
founded (weak) repairs and justified (weak) revisions correspond to justified
(weak) repairs. This result demonstrates that both formalisms, even
though rooted in different intuitions, can be ``completed'' so that to
become notational variants of each other.

Both in the case of active integrity constraints and revision programs,
the concepts of ``groundedness'' we consider do not imply, in general,
the property of the minimality of change. However, there are broad 
classes of sets of active integrity constraints, as well as classes of
revision programs when it is so. 
In the paper, we present one class of sets of active integrity constraints, for
which, independently of what database they are considered with, groundedness 
based on the notion of being justified does imply minimality, that is, for 
which justified weak repairs are minimal and so, are justified repairs (cf.
Theorem \ref{tn}). We also show that for every set of active integrity 
constraints there is a class of databases such that the minimality of 
justified weak repairs is guaranteed (cf. Theorem \ref{td}). Because of 
the correspondence between active integrity constraints and revision 
programs, one can derive analogous results for revisions programs.

A fundamental property of semantics describing database updates is the
invariance under a certain transformation of repair instances that 
consists of (1) removing some elements from a database and adding to it 
some other elements (thus, ``shifting'' the database into a different one),
and then (2) rewriting active integrity constraints by replacing literals
to reflect the changed status of some atoms in the database (cf. Section
\ref{shifting} for a detailed definition). Intuitively such a transformation,
we call it \emph{shifting}, when applied to a database and a set of integrity
constraints should result in a new database repair instance, ``isomporphic''
to the original one under any reasonable database repair semantics. We show
that it indeed is so for all the semantics we consider in the paper. Thanks 
to the correspondence between the setting of active integrity constraints 
and revision programs, the same holds true in that latter setting, too. 
Shifting is an important property. It allows us to reduce the general 
database repair (revision) problem, which is specified by two parameters, 
a database and a set of active integrity constraints (or a revision program), 
to a special case, when the database to be repaired is empty. The resulting 
setting is simpler as it involves one parameter only (a set of active 
integrity constraints or a revision program, respectively). An important
consequence of this is the existence of a direct way, in which database 
repair problem can be related to standard logic programming with the 
semantics of supported and stable models \citep{mt94c,piv01}. This paves 
the way to computational techniques for finding database repairs and 
revisions.

The paper is organized as follows. In the following
section, we situate our paper in the context of some related work. In
Section \ref{prelims}, we give a formal introduction to the database 
update problem. In Section \ref{aicIntro}, we recall basic concepts 
related to active integrity constraints, including the semantics of 
repairs and founded repairs \citep{CaroGZ}. Next, for a set of active 
integrity constraints we define weak repairs, founded weak repairs, 
justified weak repairs and justified repairs. 
We then discuss the 
\emph{normalization} of active integrity constraints in Section 
\ref{norm}. We prove that justified repairs of a database with respect
to the ``normalization'' of a set of arbitrary active integrity constraints 
are justified repairs of this database with respect to the original
(``non-normalized'') active integrity constraints (cf. Theorem \ref{tn:a}). 
This class of justified 
repairs is the most restrictive semantics for the database repair problem 
among those we consider. Thus, it offers repairs that can be regarded 
as most strongly grounded in a database repair instance (a database 
and a set of active integrity constraints).

Section \ref{complexity} 
contains complexity results concerning the existence of repairs of the
types we consider in the paper, and Section \ref{summaryRP} gives a brief 
summary of our knowledge concerning the semantics of active integrity 
constraints. In particular, we discuss there the relationships among
the semantics as well as how one could take advantage of the multitude
of the semantics considered to handle inconsistency (non-existence
of repairs of the most restrictive types).

Next, we recall basic concepts of revision programming. We then introduce 
some new semantics for revision programs. In Section \ref{connections}
we establish a precise connection between active integrity constraints and 
revision programs. We also obtain some complexity results.

Section \ref{shifting} is concerned with the shifting transformation
\citep{mt94c,piv01}. We show that all semantics discussed in the paper 
(for either formalism) are invariant under the shifting transformation 
(the proofs of those results are quite technical and we provide them in 
the appendix). The last section of the paper offers additional discussion 
of the contributions of the paper and lists some open problems. 

We close the introduction by stressing that our goal is not to single out
any of the semantics as the ``right'' one. For instance, while the semantics 
of justified repairs (revisions) seems to be best motivated by the principles 
of groundedness and minimality, the semantics given by the justified repairs 
(revisions) of the normalization of active integrity constraints (revision 
programs), being even more restrictive, certainly deserves attention. And,
in those cases when justified semantics do not offer any repairs (revisions)
relaxing the minimality requirement or the groundedness requirement
offers justified weak repairs (revisions) or founded repairs (revisions) that
one could use to enforce constraints. We discuss this matter, as well as
computational trade-offs, in Section \ref{summaryRP} and at the end of Section
\ref{summaryRev}.

\section{Related Work}

Integrity constraints may render a database inconsistent. Addressing
database inconsistency is a problem that has been studied extensively
in the literature, and several approaches to database maintenance under
integrity constraints have been proposed.

Our work is closely related to studies of \emph{event-condition-event} 
(ECA) rules in \emph{active} databases 
\citep{WidomC96c}. The main
difference is that while 
the formalisms of active integrity constraints and revision programs are
declarative, ECA rules have only been given a procedural interpretation.

To recall, an ECA rule consists of three parts:
\begin{enumerate}
\item
Event:
It specifies situations that trigger the rule (e.g. the insertion, deletion
or update of a tuple, the execution of a query, the login by a user) 
\item
Condition:
It usually models an integrity constraint. Being true in a triggered ECA
rule means the constraint is violated and causes the execution of the action
\item
Action: Typically, it is a set of update actions (\emph{insert}, 
\emph{delete}, \emph{update}). It is executed when the condition of a 
triggered rule is true.
\end{enumerate}

ECA rules without the event part are called \emph{condition-action}
(CA) rules. The structure of CA rules is similar to \emph{normal} active 
integrity constraints, as we consider them here. In this sense, the 
formalisms of ECA rules and active integrity constraints are similar.
However, there are significant differences, too. Most importantly, the 
work on ECA rules focused so far only on \emph{procedural} semantics and
particular rule processing algorithms. These algorithms determine which 
ECA rules are invoked and in what order. They use different methods for 
conflict resolution (needed when several rules are triggered at the same
time), and for ensuring termination (executing an action of a rule may 
make another triggered rule applicable, whose action in turn may make 
the first rule applicable again). 

Another approach to specify the policy for selecting a rule from 
among those that were activated was proposed by Jagadish et al. 
\citeyear{jagadish99managing}. It is based on the specification of 
a set of \emph{meta-rules} of four types:
\begin{enumerate}
\item
\emph{Positive requirement meta-rules}: A meta-rule of this type specifies
that if a rule \emph{A} executes, than a rule \emph{B} must execute as well.
\item
\emph{Disabling Rules}: A meta-rule of this type specifies if a rule \emph{A} is
executed then a rule \emph{B} will not be executed and vice versa.
\item
\emph{Preference meta-rules}: A preference meta-rule specifies a preference between two 
rules. If \emph{A} is preferable over \emph{B} and both are fireable then \emph{A} will be fired.
\item
\emph{Scheduling meta-rules}: A meta-rule of this type specifies the order 
of execution of two fireable rules. 
\end{enumerate}
Again, so far only procedural approaches to interpret meta-rules have been
developed and studied. 

In the two cases discussed, the lack of declarative semantics means there
are no grounds for a principled evaluation of rule processing algorithms. 
In contrast, in our work we focus on declarative semantics for sets of
active integrity constraints and revision programs. In particular, we 
propose several new semantics and study their properties. Our results apply
to CA rules and, in fact, they are more general, as active integrity 
constraints allow several possible actions to choose from. On the other 
hand, at present our formalisms do not allow us to specify triggering 
events.

Our work is also related to studies of \emph{consistent query answering} 
\citep{ArenasBC99,ArenasBC03}.\footnote{Chomicki \citeyear{Chomicki07}
gives an in-depth overview of this line of research.}
That research established a logical characterization of the notion of a
consistent answer in a relational database that may violate integrity 
constraints, developed properties of consistent answers, and methods to 
compute them. 

The notion of a consistent answer is based on the notion of \emph{repair}.
A repair of a database is a database that is consistent with respect to a 
given set of integrity constraints and differs minimally from the original 
one. A consistent answer to a query $Q$ over a (possibly inconsistent) 
database $\I$ with respect to a set of integrity constraints is a tuple 
that belongs to the answers to the same query over all repairs of $\I$.
Computing consistent answers exploits the notion of a \emph{residue} 
\citep{ChakravarthyGM90}. Given a query and a set of integrity constraints
over a database $\I$, instead of computing all the repairs of $\I$ and 
querying them, the consistent answers are obtained by 
computing a new query and submitting it to $\I$. The answers to the new 
query are exactly the consistent answers to the original one. The soundness, 
completeness and termination of this technique is proved for several 
classes of constraints and queries. However, the completeness is lost in 
the case of disjunctive or existential queries. Arenas, Bertossi and 
Chomicki \citeyear{ArenasBC03} present a more general approach that
allows us to compute consistent answers to any first-order query. It is
based on the notion of a logic program with exceptions. Bravo and 
Bertossi \citeyear{bravo-2006} study the problem of consistent query 
answering for databases with \emph{null} values. They propose a semantics
for integrity constraint satisfaction for that setting. Marileo and Bertossi
\citeyear{MarileoB07} developed a system for computing consistent query 
answers based on that semantics.

In research on consistent query answering, the semantics 
of choice is that of minimal change --- queries are answered with respect 
to all databases that differ minimally from the present one and that satisfy 
all integrity constraints. Thus, no distinction is made among different 
ways inconsistencies could be removed and no formalisms for specifying
policies for removing inconsistencies are discussed. The objectives
of the research on active integrity constraints and revision programs 
have been, in a sense, orthogonal. Up to now (including this paper),
the main focus was on embedding within integrity constraints declarative 
policies for removing inconsistencies, and on establishing possible 
semantics identifying candidate databases to consider as repairs.
It has not yet addressed the problem of consistent query answering with
respect to these semantics, an intriguing and important problem to address
in the future.

A closely related framework to ours was proposed and studied by Greco
et al. \citeyear{GrecoGZ03}. It was designed for computing repairs and 
consistent answers over inconsistent databases. Greco et al. 
\citeyear{GrecoGZ03} defined a repair as an inclusion-minimal set of 
update actions (insertions and deletions) that makes the database 
consistent with respect to a set of integrity constraints. The framework
relies on \emph{repair constraints}, rules that specify a set of 
insertions and deletions which are disallowed, and \emph{prioritized
constraints}, rules that define priorities among repairs. In that framework,
to compute repairs or the consistent answers, one rewrites the constraints
into a prioritized extended disjunctive logic programs with two different
forms of negation (negation as failure and classical negation). 
As shown by Caroprese et al. \citeyear{CaroGZ}, the framework can be
cast as a special case of the formalism of active integrity constraints.
A different notion of minimality, based on the cardinality of sets of 
insert and delete actions, is studied in \citep{10.1109/DEXA.2006.43}. 
This work presents a set of detailed complexity results of the problem 
of consistent query answering in the case only cardinality-based repairs 
are considered.

Katsuno and Mendelzon \citeyear{DBLP:conf/kr/KatsunoM91}, consider the 
problem of knowledge base updates. They analyze some knowledge base 
update operators and propose a set of postulates knowledge base update
operators should satisfy, but do not advocate any particular update
operator. For Katsuno and Mendelzon a knowledge base is a propositional
formula. Our setting is much more concrete as we consider databases,
knowledge bases that are conjunctions of atoms and integrity constraints
and, importantly, where updates are restricted to insertion or deletions 
of atoms. Moreover, our focus is not in update operators but on defining 
types of databases that can result from a given database when integrity 
constraints are enforced according to policies they encode. However,
the semantics we propose and study in the paper give rise to knowledge
base operators that could be considered from the standpoint of
Katsuno-Mendelzon postulates. We provide additional comments on that 
mater in the last section of the paper.

\section{Integrity Constraints and Database Repairs --- Basic Concepts}
\label{prelims}

\textbf{Databases and entailment.}
We consider a finite set $\At$ of propositional atoms. We represent 
databases as subsets of $\At$. 
A database $\I$ entails a literal $L=a$
(respectively, $L=\nt a$), denoted by $\I\models L$, if $a\in \I$ 
(respectively, 
$a\not\in\I$). Moreover, $\I$ entails a set of literals $S$, denoted
by $\I\models S$, if it entails each literal in $S$.

\noindent
\textbf{Update actions, consistency.}
Databases are \emph{updated} by 
inserting and deleting atoms. An \emph{update action} is an expression 
of the form $+a$ or $-a$, where $a\in\At$. Update action $+a$ states 
that $a$ is to be inserted. Similarly, update action $-a$ states that 
$a$ is to be deleted. We say that a set $\U$ of update actions is 
\emph{consistent} if it does not contain update actions $+a$ and $-a$, for
any $a\in\At$. 

Sets of update actions determine database updates. Let $\I$ be a 
database and $\U$ a consistent set of update actions. We define the
result of \emph{updating $\I$ by means of $\U$} as the database
\[
\I\circ\U=(\I\ \cup\ \{a\st +a\in\U\})\ \setminus\ \{a\st -a\in\U\}.
\]
We have the following straightforward property of the operator $\circ$,
which asserts that if a set of update actions is consistent, the order
in which they are executed is immaterial.

\begin{proposition}
\label{new-1017}
If $\U_1$ and $\U_2$ are sets of update actions such that $\U_1\cup\U_2$
is consistent, then for every database $\I$, $\I\circ(\U_1\cup\U_2)=
(\I\circ\U_1)\circ\U_2$
\end{proposition}

\noindent
\textbf{Integrity constraints, entailment (satisfaction).}
It is common to impose on databases conditions, called \emph{integrity
constraints}, that must always be satisfied. In the propositional 
setting, an \emph{integrity constraint} is a formula

\begin{equation}
\label{ic}
r=L_1,\ \dots,\ L_m \supset \bot,
\end{equation}
where $L_i$, $1\leq i\leq m$, are 
literals and `,' stands for the conjunction. Any subset
of $\At$ (and so, also any database) can be regarded as a propositional 
interpretation. We say that a database $\I$ \emph{satisfies} an 
integrity constraint $r$, denoted by $\I\models r$, if $\I$ satisfies 
the propositional formula represented by $r$. 
Moreover, $\I$ \emph{satisfies} a set $R$ of
integrity constraints, denoted by $\I\models R$, if $\I$ satisfies 
each integrity constraint in $R$. 
In this way, an integrity 
constraint encodes a condition on databases: the conjunction of its 
literals must not hold (or equivalently, the disjunction of the
corresponding dual literals must hold).

Any language of (propositional) logic 
could be used to describe integrity constraints (in the 
introduction we used the language with the connectives $\vee$ and 
$\neg$). Our present choice is reminiscent of the syntax used in logic 
programming. It is not coincidental.
While for integrity constraints we adopt a classical meaning of the
logical connectives, for active integrity constraints the meaning depends
on and is given by the particular semantics considered. We discuss later
several possible semantics for active integrity constraints and discuss
their properties. In most of them, the way we interpret boolean 
connectives, in particular, the negation and the disjunction, has some 
similarities to the default negation operator in logic programming and 
so, as it is common in the logic programming literature, we denote them 
with $\nt$ and $|$ rather than $\neg$ and $\vee$.

Given a set $\eta$ of integrity constraints and a database $\I$, the
problem of \emph{database repair} is to update $\I$ so that integrity
constraints in $\eta$ hold. 

\begin{definition}[\textsc{Weak Repairs and Repairs}]
\label{definition:repair}
Let $\I$ be a database and $\eta$ a set of integrity constraints. A
\emph{weak repair} for $\<\I,\eta\>$ is a consistent set $\U$ of update
actions such that $(\{+a\st a\in\I\}\cup\{-a\st a\in\At\setminus\I\})
\cap \U=\emptyset$ ($\U$ consists of ``essential'' update actions only),
and $\I\circ\U\models\eta$ (constraint enforcement).

A consistent set $\U$ of update actions is a \emph{repair} for $\<\I,
\eta\>$ if it is a weak repair for $\<\I,\eta\>$ and for every $\U'
\subseteq\U$ such that
$\I\circ \U'\models \eta$, $\U'=\U$ (minimality
of change).
~\hfill $\Box$
\end{definition}

If an original database satisfies integrity constraints (formally, if
$\I\models\eta$), then no change is needed to enforce the constraints 
and so $\U=\emptyset$ is the \emph{only} repair for $\<\I,\eta\>$. However,
there may be other \emph{weak} repairs for $\<\I,\eta\>$. This points to
the problem with weak repairs. They allow for the possibility of updating
$\I$ by means of a weak repair $\U$ for $\<\I,\eta\>$ even when $\I$ does not
violate $\eta$. Thus, the minimality of change is a natural and useful
property and, for the most part, we are interested in properties of 
repairs and their refinements. However, considering weak repairs explicitly
is useful as it offers a broader perspective.

If a set $\eta$ of integrity constraints is inconsistent, 
there is no
database satisfying it (constraints cannot be enforced). In such case, 
the database repair problem is trivial and not interesting. For that
reason, it is common in the database research to restrict investigations 
to the case when integrity constraints are consistent. However, assuming
consistency of integrity constraints does not yield any significant 
simplifications in our setting. Moreover, as we point out in the next 
section, a different notion of inconsistency arises in formalisms we study 
here that is more relevant and interesting. Therefore, in this paper, we
do not adopt the assumption that integrity constraints are consistent.

Finally, we note that the problem of the existence of a weak repair 
is NP-complete (it is just a simple reformulation of the SAT problem).
Indeed, given a database $\I$ and a set of integrity constraints
$\eta=\{L_{1,1},\dots,L_{1,m_1}\supset\bot,\dots,L_{n,1},\dots,L_{n,m_n}
\supset\bot\}$, a weak repair for $\<\I,\eta\>$ exists if and only if
$\eta$ is satisfiable (we point out that the class of propositional
integrity constraints is, modulo a standard syntactic transformation,
the same as the class of all propositional CNF theories).
Since repairs exist if and only if weak repairs do, the problem of the 
existence of a repair is NP-complete, too.

\section{Active Integrity Constraints - an Overview}
\label{aicIntro}

Given no other information but a set of integrity constraints, we have
no reason to prefer one repair over another. If several repairs are
possible, guidance on how to select a repair to execute could be useful.
The formalism of \emph{active integrity constraints} \citep{CaroGZ} was 
designed to address this problem.
We will now review it and offer a first extension by introducing
the semantics of founded weak repairs. 

\noindent
\textbf{Dual literals, dual update actions, mappings $\ua(\cdot)$ and
$\lit(\cdot)$.}
For a propositional literal $L$, we write $L^D$ for the dual literal to $L$. 
Further, if $L=a$, we define $\ua(L)=+a$. If $L=\nt a$, we define 
$\ua(L)=-a$. Conversely, for an update action $\alpha=+a$, we set 
$\lit(\alpha)=a$ and for $\alpha=-a$, $\lit(\alpha)=\nt a$. We call
$+a$ and $-a$ the \emph{duals} of each other, and write $\alpha^D$ to 
denote the update action dual to an update action $\alpha$. Finally, we 
extend the notation introduced here to sets of literals and sets of 
update actions, as appropriate.

\noindent
\textbf{Active integrity constraints, the body and head.}
An \emph{active integrity constraint} (\emph{aic}, for short) is an expression of the form

\begin{equation}\label{full-aic}
r=L_1,\ \dots,\ L_m \supset \alpha_1|\dots|\alpha_k
\end{equation}
\noindent
where $L_i$ are literals, $\alpha_j$ are update actions, and
\begin{equation}\label{updatable}
\{\lit(\alpha_1)^D,\ldots,\lit(\alpha_k)^D\}\subseteq
\{L_1,\ldots,L_m\}.
\end{equation}
The set $\{L_1,\ldots,L_m\}$ is the \emph{body} of $r$; we denote it 
by $\bd(r)$. Similarly, the set $\{\alpha_1,\ldots,\alpha_k\}$ is the 
\emph{head} of $r$; we denote it by $\hd(r)$. 

\noindent
\textbf{Active integrity constraints as integrity constraints; 
entailment (satisfaction).}
An active integrity constraint with the empty head can be regarded as
an integrity constraint (and so, we write the empty head as $\bot$, for
consistency with the notation of integrity constraints). An active integrity 
constraint with a non-empty
body can be viewed as an integrity constraint that \emph{explicitly} 
provides support for some update actions to apply. Namely, the body of 
an active integrity constraint $r$ of the form (\ref{full-aic})
represents a \emph{condition} that must be \emph{false} and so, it 
represents the integrity constraint $L_1,\ \dots,\ L_m\supset\bot$. 
Thus, we say that a database $\I$ \emph{satisfies} an active integrity 
constraint $r$ if it satisfies the corresponding integrity constraint 
$L_1,\ \dots,\ L_m\supset\bot$. We write $\I\models r$ to denote that.
This concept extends to sets of active integrity 
constraints in the standard way.
However, an active integrity constraint is more than just an integrity
constraint. It also provides support for use of update actions that are
listed in its head.

\noindent
\textbf{Updatable and non-updatable literals.}
The role of the condition (\ref{updatable}) is to ensure that an active 
integrity constraint supports only those update actions that can ``fix''
it (executing them ensures that the resulting database satisfies the 
constraint). The condition can be stated concisely as follows: 
$[\lit(\hd(r))]^D \subseteq \bd(r)$. We call literals in 
$[\lit(\hd(r))]^D$ \emph{updatable} by $r$. They are precisely those 
literals that can be affected by an update action in $\hd(r)$. We call 
every literal in $\bd(r)\setminus [\lit(\hd(r))]^D$ \emph{non-updatable}
by $r$. 
We denote the set of literals updatable by $r$
as $\upd(r)$ and the set of literals non-updatable by $r$
as $\nup(r)$.

With the notation we introduced, we can discuss the intended 
meaning of an active integrity constraint $r$ of the form (\ref{full-aic}) in more detail. 
First, $r$ functions as an integrity constraint $L_1,\dots,L_m
\supset\bot$. Second, it provides support for one of the update actions
$\alpha_i$, \emph{assuming all non-updatable literals in $r$ hold in the
repaired database}. In particular, the constraint $a,
b\supset -a | -b$, given $\I=\{a,b\}$, provides the support for $-a$ or 
$-b$, independently of the repaired database, as it has no non-updatable
literal. In the same context of $\I=\{a,b\}$, the constraint $a,b\supset
-a$ provides support for $-a$ but only if $b$ is present in the repaired
database.

It is now straightforward to adapt the concept of a (weak) repair to the 
case of active integrity constraints. Specifically, a set $\U$ of update
actions is a \emph{(weak) repair} for a database $\I$ with respect to a 
set $\eta$ of active integrity constraints if it is a repair for $\I$ 
with respect to the set of integrity constraints represented by $\eta$.

Let us consider the active integrity constraint $r=a,b \supset -b$, 
and let $\I=\{a,b\}$ be a database. Clearly, $\I$ violates $r$ as the 
condition expressed in the body of $r$ is \emph{true}. There are two
possible repairs of $\I$ with respect to $r$ or, more precisely, with
respect to the integrity constraint encoded by $r$: performing the 
update action $-a$ (deleting $a$), and performing the update action 
$-b$ (deleting $b$). 
Since $r$ provides support for the update action $-b$,
we select the latter.

Repairs do not need to obey preferences expressed by the heads of active
integrity constraints. To formalize the notion of ``support'' and
translate it into a policy to select ``preferred'' repairs, Caroprese
et al. \citeyear{CaroGZ} proposed the concept of a \emph{founded repair} --- 
a repair that is
\emph{grounded} (in some sense, \emph{implied}) by a set of active 
integrity constraints. The following definition, in addition to founded
repairs, introduces a new semantics of founded \emph{weak} repairs.

\begin{definition}[\textsc{Founded (weak) repairs}]
\label{def::actconstsem3}
Let $\I$ be a database, $\eta$ a set of active integrity constraints,
and $\U$ a consistent set of update actions.
\begin{enumerate}
\item 
An update action $\alpha$ is \emph{founded} with respect to $\<\I,\eta
\>$ and $\U$ if there is $r\in \eta$ such that $\alpha\in\hd(r)$, $\I
\circ\U\models \nup(r)$, and $\I\circ\U\models lit(\beta)^D$, for every
$\beta\in\hd(r)\setminus\{\alpha\}$.
\item
The set $\U$ is \emph{founded} with respect to $\<\I,\eta\>$ if every
element of $\U$ is founded with respect to $\<\I,\eta\>$ and $\U$.
\item
$\U$ is a \emph{founded (weak) repair} for $\<\I,\eta\>$ if $\U$ is a 
(weak) repair for $\<\I,\eta\>$ and $\U$ is \emph{founded} with respect
to $\<\I,\eta\>$. 
~\hfill $\Box$
\end{enumerate}
\end{definition}

The notion of foundedness of update actions is not restricted to update 
actions in $\U$. In other words, any update action whether in $\U$ or not 
may be founded with respect to $\<\I,\eta\>$ and $\U$. However, if an
update action, say $\alpha$, is founded with respect to $\<\I,\eta\>$ and
$\U$, and $\U$ enforces constraints, that is, $\I\circ\U \models \eta$,
then $\U$ must contain $\alpha$. Indeed, let us assume that $\alpha$ is 
founded with respect to $\<\I,\eta\>$ and $\U$ by means of an active 
integrity constraint $r\in\eta$. Let us also assume that $\I\not\models 
r$, that is, $\I\models\bd(r)$. By the foundedness, all literals in 
$\bd(r)$, except possibly for $\lit(\alpha)^D$, are satisfied in $\I
\circ\U$. Thus, since $\U$ enforces $r$, it must contain $\alpha$.
In other words, foundedness of $\alpha$ ``grounds'' $\alpha$ in $\<\I,
\eta\>$ and $\U$

In the same time, it is important to note that just foundedness of a set 
$\U$ of update actions does not imply the constraint enforcement nor the 
minimality of change. We show that in the example below. Therefore, in 
the definition of founded (weak) repairs, the property of being a (weak) 
repair must be imposed explicitly.

\begin{example}
\label{ex1}
Let $\I=\emptyset$ and
$\eta$ consist of the following active integrity constraints:
\vspace*{-0.1in}
\[
\begin{array}{llll}
r_1= & \nt a        & \supset & +a \\
r_2= & \nt b, c & \supset & +b \\
r_3= & b, \nt c  & \supset & +c.
\end{array}
\]

\noindent
The unique founded repair for $\<\I,\eta\>$ is $\{+a\}$. The set $\{+a,
+b,+c\}$ is founded, guarantees constraint enforcement (and so, it is
a founded weak repair), but it is \emph{not} change-minimal. The set
$\{+b,+c\}$ is founded but does not guarantee constraint enforcement. 
We also note that foundedness properly narrows down the class of repairs. 
If $\eta=\{a,b \supset -b\}$, and $\I=\{a, b\}$ (an example we considered 
earlier), $\U=\{-a\}$ is a repair for $\langle \I,\eta\rangle$ but not a 
founded repair.
\hfill$\Box$
\end{example}

We emphasize that founded repairs are not minimal founded weak repairs
but founded weak repairs that happen to be repairs (are minimal among all
repairs). In particular, it is possible that founded \emph{weak} repairs 
exist but founded repairs do not.
\begin{example}
\label{example:fwr_fr}
Let $\I=\emptyset$ and $\eta$ consist of the following 
active integrity constraints:
\vspace*{-0.1in}
\[
\begin{array}{llllllll}
& \nt a, b, c   & \supset & +a &\qquad\qquad 
& \nt b, a, c   & \supset & +b \\
& \nt c, a, b   & \supset & +c &\qquad\qquad
& \nt a & \supset & \bot
\end{array}
\]

\vspace*{-0.1in}
\noindent
We recall that the integrity constraint $\nt a \supset \bot$ is a special 
active integrity constraint (with an empty head).
One can check that the only founded sets of update actions are $\U_1=\emptyset$
($\emptyset$ is always vacuously founded) and $\U_2=\{+a,+b,+c\}$. Moreover,
$\U_3=\{+a\}$ is a repair and $\U_2$ is a weak repair. 
Thus, $\U_2$ is a
founded weak repair but, as it is not minimal, not a founded repair. In fact, 
there are no founded repairs in this example.
\hfill$\Box$
\end{example}

This example demonstrates that 
when we encode into 
integrity constraints a policy for selecting preferred repairs, that 
policy may be ``non-executable'' for some databases under the semantics 
of founded repairs, as founded repairs may simply not exist. Moreover,
it may be so even if the set of integrity constraints underlying the
active integrity constraints involved is consistent, that is, if weak
repairs exist (or, equivalently, if repairs exist, as repairs exist if 
and only if weak repairs do). The same is possible under the semantics 
of founded weak repairs and under all other semantics we consider later
in the paper. In other words, the assumption of consistency of integrity 
constraints does not buy us much and so, we decided not to adopt it.

Finally, we discuss the key issue arising in the context of
founded repairs that points out to the need of 
considering other semantics for active integrity constraints.
In some cases, founded 
repairs, despite combining foundedness with change-minimality, are still 
not grounded strongly enough. The problem is the circularity of support.

\begin{example}
\label{ex-923}
Let $\I=\{a,b\}$ and let $\eta_1$ consist of the following 
aic's:
\vspace*{-0.1in}
\[
\begin{array}{llll}
r_1= & a,b & \supset & -a\\
r_2= & a, \nt b   & \supset & -a \\
r_3= & \nt a, b  & \supset & -b.
\end{array}
\]

\vspace*{-0.1in}
\noindent
One can check that $\U=\{-a,-b\}$ is a repair for $\<\I,\eta_1\>$. 
Moreover, it is a founded repair: $-a$ is founded with respect 
to $\<\I,\eta_1\>$ and $\U$, with $r_2$ providing the support necessary
for foundedness of $-a$  
(i.e. Item 1 of Definition \ref{def::actconstsem3} is satisfied by 
$-a$, $\eta_1$, $\I$, $\U$ and $r_2$),
while $-b$ is founded with respect to $\<\I,\eta_1\>$ and $\U$ 
because of $r_3$ 
(i.e. Item 1 of Definition \ref{def::actconstsem3} 
is satisfied by $-b$, $\eta_1$, $\I$, $\U$ and $r_3$). 

The problem is that, arguably, $\U=\{-a,-b\}$ supports itself through
\emph{circular} dependencies. 
The constraint $r_1$ is
the only one violated by $\I$ and is the one forcing the need for a repair. 
However, $r_1$ does not support foundedness of $-a$ with respect to $\<\I,
\eta_1\>$ and $\U$, as $\I\circ\U$ does not satisfy the literal $b\in\nup(r_1)$
(required by
Item 1 of Definition \ref{def::actconstsem3}). Similarly, $r_1$ does not
support foundedness of $-b$ with respect to $\<\I,\eta_1\>$ and $\U$ (in 
fact, $-b$ is not even mentioned in the head of $r_1$). Thus, the support 
for the foundedness of $-a$ and $-b$ in $\U$ must come from $r_2$ and $r_3$
only. In fact, $r_2$ provides the support needed for 
$-a$ to be founded with respect to $\<\I,\eta_1\>$ and $\U$. However, that
requires that $b$ be absent from $\I\circ\U$ and so, $\U$ must contain the 
update action $-b$. Similarly, the support for foundedness of $-b$ is given 
by $r_3$, which requires that $a$ be absent from $\I\circ\U$, that is, that
$-a$ be in $\U$. Thus, in order for $-b$ to be founded, $\U$ must contain
$-a$, and for $-a$ to be founded, $\U$ must contain $-b$.  
In other words, the foundedness of $\{-a,-b\}$ is ``circular'': $-a$ is founded 
(and so included in $\U$) due to the fact that $-b$ has been included in
$\U$, and $-b$ is founded (and so included in $\U$) due to the fact that
$-a$ has been included in $\U$, and there is no independent justification
for having any of these two actions included --- as we noted,
$r_1$ does not ``found'' any of $-a$ nor $-b$.
~\hfill $\Box$
\end{example}

The problem of circular justifications cannot be discarded by simply hoping 
they will not occur in practice. If there are several independent sources of 
integrity constraints, such circular dependencies may arise, if only 
inadvertently.

To summarize this section, the semantics of repairs for active integrity 
constraints enforces constraints and satisfies the minimality of change
property. It has no groundedness properties beyond what is implied by 
the two requirements. The semantics of founded repairs gives preference 
to some ways of repairing constraints over others. It only considers 
repairs whose all elements are founded. However, foundedness may be 
circular and some founded (weak) repairs may be ``self-grounded'' as
in the example above. In the next section, we address the issue of
self-groundedness of founded (weak) repairs. 

On the computational side, the complexity of the semantics of repairs 
is lower than that of founded repairs. From the result stated in the
previous section, it follows that the problem of the existence of a 
repair is NP-complete, while the problem of the existence of a founded
repair is $\Sigma_P^2$-complete \citep{CaroGZ}. 
As we observed earlier,
founded repairs are not minimal founded weak repairs and, in general,
the existence of founded weak repairs is not equivalent to the existence 
of founded repairs. Thus, the complexity of the problem to decide whether 
founded weak repairs exist need not be the same as that of deciding the 
existence of founded repairs. Indeed, the complexities of the two problems 
are different (assuming no collapse of the polynomial hierarchy). Namely,
the problem of the existence of founded weak repairs is ``only'' 
NP-complete (the proof is simple and we omit it). 

\section{Justified repairs}

In this section, we will introduce another semantics for active
integrity constraints that captures a stronger concept of groundedness
than the one behind founded repairs. The goal is to disallow circular 
dependencies like the one we discussed
in Example \ref{ex-923}.

We start by defining when a set of update actions is \emph{closed} under
active integrity constraints. Let $\eta$ be a set of active integrity 
constraints and let $\U$ be a set of update actions. If $r\in\eta$,
and for every \emph{non-updatable} literal $L\in
\bd(r)$ there is an update action $\alpha\in \U$ such that $\lit(\alpha)
=L$ then, after applying $\U$ or any of its consistent supersets to the
initial database, the result of the update, say $\R$, satisfies all
non-updatable literals in $\bd(r)$. To guarantee that $\R$ satisfies 
$r$, $\R$ must \emph{falsify} at least one literal in $\bd(r)$. To 
this end $\U$ must contain at least one update action from 
$\hd(r)$. 

\noindent
\textbf{Closed sets of update actions.}
A set $\U$ of update actions is \emph{closed} under an aic
$r$ if $\nup(r)\subseteq \lit(\U)$ implies $\hd(r)\cap \U
\not=\emptyset$.  
A set $\U$ of update actions is \emph{closed} under a set $\eta$ of
active integrity constraints if it is closed under every $r\in\eta$.

If a set of update actions is not closed under a set $\eta$ of active 
integrity constraints, executing its elements does not guarantee to
enforce constraints represented by $\eta$. Therefore closed sets of
update actions are important. We regard closed sets of update actions 
that are also minimal as ``forced'' by $\eta$, as all elements in a 
minimal set of update actions closed under $\eta$ are necessary (no 
nonempty subset can be dropped).
\begin{example}
\label{ex:new:LC1}
Let us consider the database and active integrity constraints 
from Example \ref{ex-923}. The set $\mathcal{U}=\{-a,-b\}$ is
closed under $\eta_1$. We observe that the empty set is also closed 
under $\eta_1$. Therefore $\mathcal{U}$ is not minimal.
~\hfill $\Box$
\end{example}

\noindent
\textbf{No-effect actions.} Another key notion in our considerations is
that of \emph{no-effect
actions}. Let $\I$ be a database and $\R$ a result of updating $\I$. 
An update action $+a$ (respectively, $-a$) is a \emph{no-effect} action
with respect to $(\I,\R)$ if $a\in\I\cap\R$ (respectively, $a\notin\I
\cup\R$). Informally, a no-effect action does not change the status of
its underlying atom. We denote by $\nef(\I,\R)$ the set of all no-effect
actions with respect to $(\I,\R)$. We note the following two simple
properties reflecting the nature of no-effect actions --- their 
redundancy.

\begin{proposition}
\label{new-930a}
Let $\I$ be a database. Then
\begin{enumerate}
\item For every databases $\R,\R'$, if $\nef(\I,\R)\subseteq
\nef(\I,\R')$, then $\R'\circ\nef(\I,\R)=\R'$
\item For every set $\E$ of update actions such that $\E\cup\nef(\I,
\I\circ \E)$ is consistent, if $\E'\subseteq\E$, then 
$\I\circ\E'=\I\circ(\E'\cup\nef(\I,\I\circ\E))$.
\end{enumerate}
\end{proposition}
\textbf{Proof:}
\begin{enumerate}
\item
Since $\nef(\I,\R)=\{+a\st a\in\I\cap\R\}\cup\{-a\st a\notin\I\cup\R\}$
and $\nef(\I,\R')=\{+a\st a\in\I\cap\R'\}\cup\{-a\st a\notin\I\cup\R'\}$,
we have $\I\cap\R\subseteq\I\cap\R'$ and $\I\cup\R'\subseteq\I\cup\R$.
It follows that $\I\cap\R\subseteq\R'$ and $\R'\subseteq\I\cup\R$. Thus,
$\R'\circ\nef(\I,\R)=(\R'\cup(\I\cap\R))\cap(\I\cup\R)=\R'$.
\item As $\E'\subseteq\E$, then $\nef(\I,\I\circ\E)\subseteq
\nef(\I,\I\circ\E')$.
Since $\E\cup\nef(\I,\I\circ\E)$ is consistent, Propositions
\ref{new-1017} and \ref{new-930a}(1) imply that $\I\circ(\E'\cup\nef(\I,\I
\circ\E))=(\I\circ\E')\circ\nef(\I,\I\circ\E)=\I\circ\E'$.
~\hfill$\Box$
\end{enumerate}

\smallskip
Our semantics of justified repairs is based
on the knowledge-representation principle, a form of the frame
axiom \citep{mh69}, that remaining in the previous state requires no reason 
(persistence by inertia). Thus, when justifying update actions necessary
to transform $\I$ into $\R$ based on $\eta$ we assume the set $\nef(\I,
\R)$ as given. This brings us to the notion of a justified weak repair.

\begin{definition}[\textsc{Justified weak repairs}]
\label{ejc}
Let $\I$ be a database and $\eta$ a set of active integrity 
constraints. A consistent set $\U$ of update actions is a justified 
action set for $\<\I,\eta\>$ if $\U$ is a minimal 
set of update actions containing $\nef(\I,\I\circ\U)$ and closed 
under $\eta$.

If $\U$ is a justified action set for $\<\I,\eta\>$, then $\E=\U\setminus
\nef(\I,\I\circ\U)$ is a justified weak repair for $\<\I,\eta\>$.
~\hfill$\Box$
\end{definition}

Intuitively, a set $\U$ of update actions is a justified action set, if
it is precisely the set of update actions forced or \emph{justified} by 
$\eta$ and the no-effect actions with respect to $\I$ and $\I\circ\U$.
This ``fixpoint'' aspect of the definition is reminiscent of 
the definitions of semantics of several non-monotonic logics, including 
(disjunctive) logic programming with the answer set semantics. The
connection can be made more formal and we take advantage of it in the
section on the complexity and computation.

Before we proceed, we will illustrate the notion of justified weak repairs. 

\begin{example}
\label{ex-925new} 
Let us consider again Example \ref{ex-923}. 
The set $\mathcal{U}=\{-a,-b\}$ is not a justified weak repair for 
$\<\I,\eta_1 \>$. One can check that $\mathcal{U}\cup\nef(\I,\I\circ
\mathcal{U})$ ($=\{-a,-b\}$) contains $\nef(\I,\I\circ\mathcal{U})$ 
($=\emptyset$), and 
is closed under $\eta_1$. But, as we observed in Example \ref{ex:new:LC1},
it is not a minimal set of update actions containing $\nef(\I,\I\circ
\mathcal{U})$ and closed under $\eta_1$. Indeed, $\emptyset$ has these 
two properties, too. In fact, one can check that $\<\I,\eta_1\>$ has no 
justified weak repairs.

Next, let us consider a new set, $\eta_2$, of aic's, where $r_1$ is replaced 
with $r'_1=a,b \supset -a|-b$.
The constraint $r'_1$ provides support for $-a$ or $-b$ independently 
of the repaired database (as there are no non-updatable literals in $r'_1$).
If $-a$ is selected (with support from $r'_1$), $r_3$ supports $-b$. If 
$-b$ is selected (with support from $r'_1$), $r_2$ supports $-a$, Thus the
cyclic support given by $r_2$ and $r_3$ in the presence of $r_1$ is 
broken. Indeed, one can check that $\{-a,-b\}$ is a justified weak repair,
in fact, the only one.
~\hfill $\Box$
\end{example}

We note that the set $\nef(\I,\I\circ\U)$ can be quite large. In particular,
the cardinality of the set of update actions $-a$, where $a\not\in \I\cup 
\R$, cannot be bounded by the size of the database repair problem, which
is given by the size of $\I$ and $\eta$. However, only those update actions
$-a$ of that type are important from the perspective of justified weak 
revisions, whose literals $\nt(a)$ occur in the bodies of active integrity 
constraints in $\eta$ (as no other update action of that type can play a 
role in determining minimal sets of update actions closed under integrity 
constraints). 

We will now study justified action sets and justified weak repairs. 
We start with an alternative characterization of justified weak repairs.

\begin{theorem}
\label{cor-930}
Let $\I$ be a database, $\eta$ a set of active integrity constraints
and $\E$ a consistent set of update actions. Then $\E$ is 
a justified weak repair for $\<\I,\eta\>$ if and only if
$\E\cap \nef(\I,\I\circ\E)=\emptyset$ and $\E\cup\nef(\I,\I\circ\E)$
is a justified action set for $\<\I,\eta\>$.
\end{theorem}
\textbf{Proof:} ($\Rightarrow$) Since $\E$ is a justified weak repair
for $\<\I,\eta\>$, $\E=\U\setminus\nef(\I,\I\circ\U)$ for some 
consistent set $\U$ of update actions such that $\U$ is minimal 
containing $\nef(\I,\I\circ\U)$ and closed under $\eta$. By Proposition
\ref{new-930a}(2), $\I\circ\U=\I\circ\E$. Thus, $\E\cap\nef(\I,\I\circ
\E)=\emptyset$. Moreover, since $\nef(\I,\I\circ\U)\subseteq\U$, $\U=
\E\cup\nef(\I,\I\circ\E)$. Hence, $\E\cup\nef(\I,\I\circ\E)$ is a 
justified action set for $\<\I,\eta\>$.

\smallskip
\noindent
($\Leftarrow$) Let $\U=\E\cup\nef(\I,\I\circ\E)$. We will show that
$\nef(\I,\I\circ\U)=\nef(\I,\I\circ\E)$. To this end, let $+a\in\nef(\I,
\I\circ\U)$. Then, $a\in\I$ and $-a\notin\U$ (the latter property 
follows by the consistency of $\U$). It follows that $-a\notin\E$
and, consequently, $+a\in\nef(\I,\I\circ\E)$. Similarly, we show that
if $-a\in\nef(\I,\I\circ\U)$, then $-a\in\nef(\I,\I\circ\E)$. Thus, we
obtain that $\nef(\I,\I\circ\U)\subseteq\nef(\I,\I\circ\E)$.

Conversely, let $+a\in\nef(\I,\I\circ\E)$. Then $a\in\I$ and $+a\in\U$. 
Since $\U$ is consistent (it is a justified action set for
$\<\I,\eta\>$), $\I\circ\U$ is well defined and $+a\in\nef(\I, \I\circ\U)$. 
The case $-a\in\nef(\I,\I\circ\E)$ is similar. Thus, $\nef(\I,\I\circ\E)
\subseteq\nef(\I,\I\circ\U)$ and the claim follows.

Since $\E\cap\nef(\I,\I\circ\E)=\emptyset$, we obtain that $\E=\U
\setminus\nef(\I,\I\circ\U)$. Since $\U$ is a justified action
set for $\<\I,\eta\>$, $\E$ is a justified weak repair for $\<\I,\eta\>$.
~\hfill$\Box$

\smallskip
Justified weak repairs have two key properties for the problem of
database update: constraint enforcement (hence the term ``weak repair'')
and foundedness.

\begin{theorem}
\label{prop1}
Let $\I$ be a database, $\eta$ a set of active integrity constraints,
and $\E$ a justified weak repair for $\<\I,\eta\>$. Then
\begin{enumerate}
\item For every atom $a$, exactly one of $+a$ and $-a$ is in
$\E\cup\nef(\I,\I\circ\E)$
\item $\I\circ\E\models\eta$
\item $\E$ is founded for $\<\I,\eta\>$.
\end{enumerate}
\end{theorem}
\textbf{Proof:} 
Throughout the proof, use the notation $\U=\E\cup
\nef(\I,\I\circ\E)$. 

\begin{enumerate}
\item Since $\U$ is consistent (cf. Theorem \ref{cor-930}), for every 
atom $a$, at most one of $+a$, $-a$ is in $\U$. If $+a\in\nef(\I,\I
\circ\E)$ or $-a\in\nef(\I,\I\circ\E)$ then the claim follows. 
Otherwise, the status of $a$ changes as we move from $\I$ to $\I\circ
\E$. That is, either $+a$ or $-a$ belongs to $\E$ and, consequently, 
to $\U$, as well.

\item Let us consider $r\in\eta$. Since $\U$ is closed under $\eta$ (cf.
Theorem \ref{cor-930}), we have $\nup(r)\not\subseteq\lit(\E\cup\nef(\I,
\I\circ\E))$ or $\hd(r)\cap(\E\cup\nef(\I,\I\circ\E))\not=\emptyset$.
Let us assume the first possibility, and let $L$ be a literal such that
$L\in\nup(r)$ and $\ua(L)\notin\U$. By (1), $\ua(L^D)\in\U$. 
Consequently, $\I\circ\U\not\models L$. By Proposition \ref{new-930a}(2),
$\I\circ\E\not\models L$. Since $L\in\bd(r)$, $\I\circ\E\models r$.

Thus, let us assume that $\hd(r)\cap\U\not=\emptyset$ and
let $\alpha\in\hd(r)\cap\U$.
Then $\alpha\in\hd(r)$ and so, $\lit(\alpha)^D \in\bd(r)$. Furthermore,
$\alpha\in\U$ and so, $\I\circ\U\models\lit(\alpha)$. By Proposition
\ref{new-930a}(2), $\I\circ
\E\models\lit(\alpha)$. Thus, $\I\circ\E\models r$ in this case, 
too.

\item Let $\alpha\in\E$. By Theorem \ref{cor-930}, $\alpha\notin
\nef(\I,\I\circ\E)$. Thus, $\nef(\I,\I\circ\E)\subseteq\U\setminus\{
\alpha\}$. Since $\U$ is a minimal set closed under $\eta$ and 
containing $\nef(\I,\I\circ\E)$, $\U\setminus\{\alpha\}$ is not
closed under $\eta$. That is, there is $r\in\eta$ such that 
$\nup(r)\subseteq\lit(\U\setminus\{\alpha\})$ and $\hd
(r)\cap(\U\setminus\{\alpha\})=\emptyset$.
We have
\begin{eqnarray*}
\I\circ(\U\setminus\{\alpha\}) &=& \I\circ(\nef(\I,\I\circ\E)\cup(\E
\setminus\{\alpha\}))\\
&=& (\I\circ\nef(\I,\I\circ\E))\circ(\E\setminus
\{\alpha\}).
\end{eqnarray*}
By Proposition \ref{new-930a} (and the fact that $\nef(\I,\R)=\nef(\R,
\I)$, for every databases $\I$ and $\R$),
\begin{equation}
\label{eq-1016}
\I\circ(\U\setminus\{\alpha\})=\I\circ(\E\setminus\{\alpha\}).
\end{equation}
From $\nup(r)\subseteq\lit(\U\setminus\{\alpha\})$, it follows that
$\I\circ(\U\setminus\{\alpha\})\models\nup(r)$. By (\ref{eq-1016}),
$\I\circ(\E\setminus\{\alpha\})\models\nup(r)$. Since $\alpha\in\hd(r)$,
$\lit(\alpha^D)\notin\nup(r)$. Thus, $I\circ\E\models \nup(r)$.

The inclusion $\nup(r)\subseteq\lit(\U\setminus\{\alpha\})$ also implies
$\nup(r)\subseteq \lit(\U)$. Since $\U$ is closed under $\eta$, $\hd(r)
\cap\U\not=\emptyset$ and so, $\hd(r)\cap\U=\{\alpha\}$.

Let us consider $\beta\in\hd(r)$ such that $\beta\not=\alpha$. It follows
that $\beta\notin\U$. By (1), $\beta^D\in \U$ and, consequently, $I\circ
\U\models\beta^D$. Since $\I\circ\U=\I\circ\E$ (Proposition 
\ref{new-930a}), it follows that $\alpha$
is founded with respect to $\<\I,\eta\>$ and $\E$.
~\hfill$\Box$
\end{enumerate}

Theorem \ref{prop1} directly implies that justified weak repairs are 
founded weak repairs.

\begin{corollary}
\label{cor-925}
Let $\I$ be a database, $\eta$ a set of active integrity constraints,
and $\E$ a justified weak repair for $\<\I,\eta\>$. Then, $\E$ is a 
founded weak repair for $\<\I,\eta\>$. 
\end{corollary}

Examples \ref{ex-923} and \ref{ex-925new} show that the converse to 
Corollary \ref{cor-925} does not hold. That is, there are
founded weak repairs that are not justified weak repairs.

While a stronger property than foundedness, being a justified weak repair
still does not guarantee change-minimality (and so, the term
\emph{weak} cannot be dropped).

\begin{example}
\label{ex-925a}
Let $\I'=\emptyset$, and $\eta_3$ be a set of aic's consisting of

\vspace*{-0.1in}
\[
\begin{array}{llll}
r_1= & \nt a, b    & \supset & +a|-b\\
r_2= & a, \nt b          & \supset & -a|+b.
\end{array}
\]
Clearly, $\I'$ is consistent with respect to $\eta_3$.
Let us consider the set of update actions $\E=\{+a,+b\}$. It is easy to 
verify that $\E$ is a justified weak repair for $\<\I',\eta_3\>$.
Therefore, it ensures constraint enforcement and it is founded.
However, $\E$ is not minimal and the empty
set of update actions is its only repair. 
~\hfill$\Box$
\end{example}

Thus, to have change-minimality, it needs to be enforced directly as
in the case of founded repairs. By doing so, we obtain the notion
of \emph{justified repairs}. 

\begin{definition}[\textsc{Justified repair}]
\label{ejr}
Let $\I$ be a database and $\eta$ a set of active integrity constraints.
A set $\E$ of update actions is a \emph{justified repair} for $\<\I,
\eta\>$ if $\E$ is a justified weak repair for $\<\I,\eta\>$, and for every $\E' \subseteq\E$
such that $\I\circ\E'\models\eta$, $\E'=\E$.
~\hfill$\Box$
\end{definition}

Theorem \ref{prop1} has yet another corollary, this time concerning 
justified and founded repairs.
\begin{corollary}
\label{jsdisfdd}
Let $\I$ be a database, $\eta$ a set of active integrity constraints,
and $\E$ a justified repair for $\<\I,\eta\>$. Then, $\E$ is a 
founded repair for $\<\I,\eta\>$. 
\end{corollary}
\textbf{Proof:} Let $\E$ be a justified repair for $\<\I,\eta\>$. It 
follows by Theorem \ref{prop1} that $\I\circ\E\models\eta$. Moreover,
by the definition
of justified repairs, $\E$ is change minimal. Thus, $\E$ is a repair.
Again by Theorem \ref{prop1}, $\E$ is founded. Thus, $\E$ is a founded
repair for $\<\I,\eta\>$.~\hfill $\Box$

Examples \ref{ex-923} and \ref{ex-925new} show that the inclusion 
asserted by Corollary \ref{jsdisfdd} is proper. Indeed, we argued in
Example \ref{ex-923} that $\{-a,-b\}$ is a founded repair. Then, in
Example \ref{ex-925new} we showed that it is not a justified weak repair.
Thus, $\{-a,-b\}$ is not a justified repair, either.

As illustrated by Example \ref{ex-925a}, in general, justified  
repairs form a proper subclass of justified weak repairs. However, in some 
cases the two concepts coincide --- the minimality is a consequence of
the groundedness underlying the notion of a justified weak repair. 
One such case is identified in the next theorem. The other important
case is discussed in the next section.
\begin{theorem}
\label{td}
Let $\I$ be a database and $\eta$ a set of active integrity constraints
such that for each update action $\alpha \in \bigcup_{r\in \eta}head(r)$,
$\I \models lit(\alpha^D)$. If $\E$ is a justified weak repair
for $\<\I,\eta\>$, then $\E$ is a justified repair for $\<\I,\eta\>$.
\end{theorem}
\textbf{Proof:}
Let $\E$ be a justified weak repair for $\<\I,\eta\>$ and let $\E'
\subseteq\E$ be such that $\I\circ\E'\models\eta$.

We define $\U=\E\cup\nef(\I,\I\circ\E)$. By Theorem \ref{cor-930} and
Proposition \ref{new-930a}(2), $\U$ is a minimal set of update actions
containing
$\nef(\I,\I\circ\E)$ and closed under $\eta$. Let $\U'=\E'\cup\nef(\I,
\I\circ\E)$ and let $r\in\eta$ be such that
$\ua(\nup(r))\subseteq\U'$. Since $\I\circ\E'\models\eta$, $\I\circ\E'
\not\models\bd(r)$. Thus, it follows that there is $L\in\upd(r)$ such
that $\I\circ\E'\not\models L$. Since $L\in\upd(r)$, there is $\alpha
\in \hd(r)$ such that $L=\lit(\alpha^D)$. By the assumption, $\I\models
L$, that is, $\I\models \lit(\alpha^D)$. Since $\I\circ\E'\not\models 
L$, $\I\circ\E'\models\lit(\alpha)$. Thus, $\alpha\in\E'$ and, 
consequently, $\alpha\in\U'$. It follows that $\U'$ is closed under $r$
and, since $r$ was an arbitrary element of $\eta$, under $\eta$. too.
Thus, $\U'=\U$, that is, $\E'\cup\nef(\I,\I\circ\E)=\E\cup\nef(\I,\I\circ
\E)$. Since $E'\subseteq\E$ and $\E\cap\nef(\I,\I\circ\E)=\emptyset$, 
$\E'=\E$. It follows that $\E$ is a minimal set of update actions such 
that $\I\circ\E\models\eta$.
~\hfill $\Box$

The theorem above states that whenever each update action occurring in 
$\eta$ is essential with respect to $\I$ 
(it is able to perform a real change over $\I$), the minimality of each
justified weak repair is guaranteed (that is, it is a justified repair).

\section{Normal active integrity constraints and normalization}
\label{norm}

An active integrity constraint $r$ is \emph{normal} if 
$|\hd(r)|\leq 1$.
We will now study properties of normal active integrity constraints. 
First, we will show that for that class of constraints, updating by
justified weak repairs guarantees the minimality of change
property and so, the explicit reference to the latter can be omitted from
 the definition of justified repairs.

\begin{theorem}
\label{tn}
Let $\I$ be a database and $\eta$ a set of normal active integrity
constraints. If $\E$ is a justified weak repair for $\<\I,\eta\>$
then $\E$ is a justified repair for $\<\I,\eta\>$.
\end{theorem}
\textbf{Proof:}
Let $\E$ be a justified weak repair for $\<\I,\eta\>$. We have to prove
that $\E$ is minimal with respect to constraint enforcement. To this
end, let us consider $\E'\subseteq\E$ such that $\I\circ\E'\models\eta$.

We define $\U=\E\cup\nef(\I,\I\circ\E)$ and $\U'=\E'\cup\nef(\I,\I\circ
\E)$. We will show that $\U'$ is closed under $\eta$. Let $r\in\eta$ be
such that $\ua(\nup(r))\subseteq\U'$.

Since $\I\circ\E'\models r$, $\I\circ\E'\not\models \bd(r)$. By our
assumption, $\ua(\nup(r))\subseteq\U'$. Thus, $\I\circ\U'\models\nup(r)$.
Since $\U'$ is consistent, Proposition \ref{new-930a}(2) implies that
$\I\circ\E'=\I\circ\U'$. Thus, $\I\circ\E'\models\nup(r)$. If $\hd(r)=
\emptyset$, $\I\circ\E'\models\bd(r)$ and so, $\I\circ\E'\not\models r$,
a contradiction. Thus, $\hd(r)=\{\alpha\}$, for some update action
$\alpha$. Moreover, as $\I\circ\E'\models r$, $\I\circ\E'\not\models
\lit(\alpha^D)$. Consequently, $\I\circ\E'\models\lit(\alpha)$.

Since $\U'\subseteq\U$, $\ua(\nup(r))\subseteq\U$. By Theorem 
\ref{cor-930}, $\U$ is closed under $\eta$. Thus, $\alpha\in\U$. Since
$\I\circ\U=\I\circ\E$ (Proposition \ref{new-930a}(2)),
$\I\circ\E\models \lit(\alpha)$.

If $\I\models\lit(\alpha)$ then, as $\I\circ\E\models \lit(\alpha)$,
we have $\alpha\in\nef(\I,\I\circ\E)\subseteq\U'$. If $\I\not\models 
\lit(\alpha)$ then, as $\I\circ\E'\models\lit(\alpha)$, we have that
$\alpha\in\E'\subseteq\U'$. Thus, $\U'$ is closed under $r$ and so, 
also under $\eta$. Consequently, $\U'=\U$. Since $\E\cap\nef(\I,\I\circ
\E)=\emptyset$, it follows that $\E'=\E$. Thus, $\E$ is a minimal set 
of update actions such that $\I\circ\E\models\eta$.
~\hfill $\Box$

\noindent
\textbf{Normalization.}
Next, we introduce the operation of \emph{normalization} of active
integrity constraints, which consists of eliminating disjunctions from
the heads of rules. For an active integrity constraint
$
r=\phi\supset \alpha_1|\dots|\alpha_n,
$
by $r^n$ we denote the set of \emph{normal} active integrity constraints 
$
\{
\phi\supset \alpha_1,\dots,\phi\supset\alpha_n
\}
$. For a set $\eta$ of active integrity constraints, we set
$\eta^n=\bigcup_{r\in \eta} r^n$. It is shown by Caroprese et al. 
\citeyear{CaroGZ} that
$\E$ is founded for $\<\I,\eta\>$ if and only if $\E$ is a
founded for $\<\I,\eta^n\>$. Thus, $\E$ is a founded (weak) repair for
$\<\I,\eta\>$ if and only if $\E$ is a founded (weak) repair for
$\<\I,\eta^n\>$. For justified repairs, we have a weaker result. 
Normalization may eliminate some justified repairs. That leads to an
even more narrow class of repairs than justified ones, an issue we discuss
later in Section \ref{summaryRP}.

\begin{theorem}
\label{tn:a}
Let $\I$ be a database and $\eta$ a set of active integrity constraints.
\begin{enumerate}
\item If a set $\E$ of update actions is a justified repair for $\<\I,
\eta^n\>$, then $\E$ is a justified repair for $\<\I,\eta\>$
\item If a set $\E$ of update action is a justified weak repair for $\<\I,
\eta^n\>$, then $\E$ is a justified weak repair for $\<\I,\eta\>$.
\end{enumerate}
\end{theorem}
\textbf{Proof:}
Let $\E$ be a justified repair for $\<\I,\eta^n\>$. We define $\U=\E\cup
\nef(\I,\I\circ\E)$. By Corollary \ref{jsdisfdd}, $\E$ is a founded 
repair for $\<\I,\eta^n\>$. By a result obtained by Caroprese et al.
\citeyear{CaroGZ}, $\E$ is a 
founded repair for $\<\I,\eta\>$ and, consequently, a repair for $\<\I,
\eta\>$.

Since $\E$ is, in particular, a justified weak repair for $\<\I,
\eta^n\>$, $\U$ is a justified action set for $\<\I,\eta^n\>$ (Theorem 
\ref{cor-930}). Thus, $\U$ is a minimal set of update actions containing
$\nef(\I,\I\circ\E)$ and closed under $\eta^n$. To prove that $\E$ is a 
justified repair for $\<\I,\eta\>$, it suffices to show that $\U$ is 
a minimal set of update actions containing $\nef(\I,\I\circ\E)$ and 
closed under $\eta$.

Let us consider an active integrity constraint
$$
r=lit(\alpha_1^D),\dots,lit(\alpha_n^D),\phi\supset \alpha_1|\dots|
\alpha_n
$$
in $\eta$ such that $\ua(\nup(r))\subseteq\U$
(we note that $\nup(r)$ consists precisely of the literals that appear 
in $\phi$). It follows that $I\circ\U\models\nup(r)$. Since $\E$ is a 
repair, $\I\circ\E\not\models\bd(r)$. By Proposition \ref{new-930a}(2),
$\I\circ\E=\I\circ\U$. Thus, $\I\circ\U\not\models\bd(r)$. It follows
that there is $i$, $1\leq i\leq n$, such that $\I\circ\U\not\models\lit
(\alpha_i^D)$. Thus, $\alpha_i^D\notin\U$. By Theorem \ref{prop1}(1), 
$\alpha_i\in\U$. Thus, $\U$ is closed under $r$ and, consequently, under
$\eta$, as well.

We will now show that $\U$ is minimal in the class of sets of update 
actions containing $\nef(\I,\I\circ\E)$ and closed under $\eta$.  
Let $\U'$ be a set of update actions such that $ne(\I,\I\circ\E)
\subseteq\U'\subseteq \U$ and $\U'$ is closed under $\eta$. Let us
consider an active integrity constraint in $s\in\eta^n$ such that 
$\ua(\nup(s))\subseteq \U'$. 

By the definition of $\eta^n$, there is an active integrity constraint
$r\in\eta$ such that
$$
r=lit(\alpha_1^D),\dots,lit(\alpha_i^D),\dots,lit(\alpha_n^D),\phi
\supset \alpha_1|\dots|\alpha_i|\dots|\alpha_n
$$
and
$$
s=lit(\alpha_1^D),\dots,lit(\alpha_i^D),\dots,lit(\alpha_n^D),\phi\supset
\alpha_i.
$$
Since $\ua(\nup(s)\subseteq\U'$, $\ua(\nup(r)\subseteq\U'$.
As $\U'$ 
is closed under $\eta$, there is $j$, $1\leq j\leq n$, such that 
$\alpha_j\in\U'$. For every $k$ such that $1\leq k\leq n$ and $k\not=
i$, $\alpha_k^D\in\U'$. By the consistency of $\U'$, we conclude that 
$\alpha_i\in \U'$. Thus, $\U'$ is closed under $s$ and, consequently,
under $\eta^n$. Since $\U'\subseteq\U$ and $\U$ is minimal containing 
$ne(\I,\I\circ\E)$ and closed under $\eta^n$ it follows that $\U'=\U$.
Thus, $\U$ is minimal containing $ne(\I,\I\circ\E)$ and closed under 
$\eta$. Consequently, $\E$ is a justified repair for $\<\I,\eta\>$.

\smallskip
\noindent
(2) If $\E$ is a justified weak repair for $\<\I,\eta^n\>$ then, by 
Theorem \ref{tn}, $\E$ is a justified repair for $\<\I,\eta^n\>$. By
(1), $\E$ is a justified repair for $\<\I,\eta\>$ and so, a justified
weak repair for $\<\I,\eta\>$. 
~\hfill $\Box$

The following example shows that the inclusions in the previous theorem
are, in general, proper.
\begin{example}
\label{ex-925b}
Let us consider an empty database $\I'=\emptyset$, the set $\eta_4$ of
aic's

\vspace*{-0.1in}
\[
\begin{array}{llll}
 r_1=&\nt a, \nt b    & \supset & +a|+b\\
 r_2=&a, \nt b        & \supset & +b\\
 r_3=&\nt a,  b       & \supset & +a,
\end{array}
\]
its normalized version $\eta_4^n$

\vspace*{-0.1in}
\[
\begin{array}{lllllllll}
 r_{1,1}=&\nt a, \nt b    & \supset & +a &\qquad &
  r_{2,1}=&a, \nt b        & \supset & +b\\
 r_{1,2}=&\nt a, \nt b    & \supset & +b &\qquad &
 r_{3,1}=&\nt a,  b       & \supset & +a,
\end{array}
\]
and the set of update actions $\E=\{+a,+b\}$. It is easy to verify that
$\E$ is a justified repair for $\<\I',\eta_4\>$. However, $\E$ is not a 
justified weak repair for $\<\I',\eta_4^n\>$ (and so, not a justified
repair for $\<\I',\eta_4^n\>$). Indeed, it is not a minimal set 
containing $\nef(\I',\I'\circ\E)=\emptyset$ and closed under $\eta_4^n$,
as $\emptyset$ is also closed under $\eta_4^n$. ~\hfill$\Box$
\end{example}

\section{Complexity and Computation}
\label{complexity}
We noted earlier that the problem of the existence of a (weak) repair
is NP-complete, and the same is true for the problem of the existence of
founded weak repairs. On the other hand, the problem of the existence
of a founded repair is $\Sigma_P^2$-complete \citep{CaroGZ}. In this
section, we study the problem of the existence of justified (weak)
repairs.

For our hardness results, we will use problems in logic programming.
We will consider disjunctive and normal logic programs that satisfy 
some additional syntactic constraints. Namely, we will consider only
programs without rules which contain multiple occurrences of the same 
atom (that is, in the head and in the body, negated or not; or in the
body --- both positively and negatively). We call such programs 
\emph{simple}. It is well known that the problem of the existence of
a stable model of a normal logic program is NP-complete \citep{mt88}, and
of the disjunctive logic program --- $\Sigma_2^P$-complete \citep{eg95}.
The proofs provided by Marek and Truszczy\'{n}ski
\citeyear{mt88} and Eiter and Gottlob \citeyear{eg95} imply that the results hold also under 
the restriction to simple normal and simple disjunctive programs, 
respectively (in the case of disjunctive logic programs, a minor 
modification of the construction is required).

Let $\rho$ be a logic programming rule, say
\[
\rho = a_1|\ldots|a_k\leftarrow\beta.
\]
We define 
\[
aic(\rho) = \nt a_1,\dots,\nt a_k,\beta\supset +a_1|\dots|+a_k .
\]
We extend the operator $aic(\cdot)$ to logic programs in a standard
way. We note that if a rule $\rho$ is simple, then $\bd(aic(\rho))$
is consistent and $\nup(aic(\rho))=\bd(\rho)$.

We recall that a set $M$ of atoms is an answer set of a disjunctive
logic program $P$ if $M$ is a minimal set closed under the reduct $P^M$,
where $P^M$ consists of the rules obtained by dropping all negative 
literals from those rules in $P$ that do not contain a literal $\nt a$
in the body, for any $a\in M$ (we refer to the paper by Gelfond and
Lifschitz \citeyear{gl90b} for details).
Our first two lemmas establish a result needed for hardness arguments.

\begin{lemma}
\label{new-1017c}
Let $P$ be a simple disjunctive logic program and $M',M$ sets of atoms
such that $M'\subseteq M$. Then $M'$ is a model of $P^M$ if and only if
$\{+a\st a\in M'\}\cup\{-a\st a\notin M\}$ is closed under $aic(P)$.
\end{lemma}
\textbf{Proof:}  Let us define $\U=\{+a\st a\in M'\}\cup\{-a\st a\notin
M\}$. We note that $\U$ is consistent.

\smallskip
\noindent
($\Rightarrow$) Let $r\in aic(P)$, $\rho\in P$ be a rule such
that $r=aic(\rho)$, and $\rho'$ be the rule obtained by eliminating 
from $\rho$ all negative literals.

Since $P$ is simple, $\nup(r)=\bd(\rho)$. Let 
us assume that $\nup(r)\subseteq\U$. It follows that $\rho'\in P^M$ and
that $M'\models\bd(\rho')$. Thus, $\hd(\rho')\cap M'\not=\emptyset$.
Since $\hd(\rho)=\hd(\rho')$ and $\hd(r)=\hd(aic(\rho))=\ua(\hd(\rho))$, 
$\hd(r)\cap\U\not=\emptyset$. That is, $\U$ is closed under $r$ and,
since $r$ was chosen arbitrarily, under $aic(P)$, too.

\smallskip
\noindent
($\Leftarrow$) Let us consider $\rho'\in P^M$. There is $\rho\in P$
such that for every negative literal $\nt a \in \bd(\rho)$, 
$a\notin M$, and dropping all negative literals from $\rho$ results in
$\rho'$. If $\bd(\rho')\subseteq M'$, then $\bd(\rho)\subseteq\lit(\U
)$. Thus, $\nup(aic(\rho))\subseteq\U$. It follows that
$\hd(aic(\rho))\cap\U \not=\emptyset$. 
Thus, $\hd(\rho)\cap\lit(\U)\not=\emptyset$. Since $\hd(\rho)$ consists
of atoms and $\hd(\rho')=\hd(\rho)$, $\hd(\rho')\cap M'\not=\emptyset$.
That is, $M'\models \rho'$ and, consequently, $M'\models P^M$.
~\hfill$\Box$

\begin{theorem}
\label{new-1017b}
Let $P$ be a simple disjunctive logic program. A set $M$ of atoms is
an answer set of $P$ if and only if $\ua(M)$ is a justified weak repair
for $\<\emptyset,aic(P)\>$.
\end{theorem}
\textbf{Proof:} ($\Rightarrow$) Let $M$ be an answer set of $P$. That 
is, $M$ is a minimal set closed under the rules in the reduct $P^M$. 
By Lemma \ref{new-1017c}, $\{+a\st a\in M\}\cup\{-a\st a\notin M\}$
is closed under $aic(P)$. Let $\U'$ be a set of update actions such
that $\{-a\st a\notin M\}\subseteq\U'\subseteq \{+a\st a\in M\}\cup\{-a
\st a\notin M\}$. We define $M'=\{a\st +a\in\U'\}$. Then $M'\subseteq
M$. By Lemma \ref{new-1017c}, $M'\models P^M$. Since $M$ is an answer
set of $P$, $M'=M$ and $\U'=\U$. It follows that $\{+a\st a\in M\}\cup
\{-a\st a\notin M\}$ is a minimal set closed under $aic(P)$ and
containing $\{-a\st a\notin M\}$. Since $\ua(M)=\{+a\st a\in M\}$ and
$\nef(\emptyset,\emptyset\circ \ua(M))=\{-a\st a\notin M\}$, Theorem 
\ref{cor-930} implies that $\ua(M)$ is justified weak repair for 
$\<\emptyset,aic(P)\>$.

\smallskip
\noindent
($\Leftarrow$) By Theorem \ref{cor-930}, $\{+a\st a\in M\}\cup\{-a\st 
a\notin M\}$ is a minimal set containing $\{-a\st a\notin M\}$ and
closed under $aic(P)$. By Lemma \ref{new-1017c}, $M$ is a model of 
$P^M$. Let $M'\subseteq M$ be a model of $P^M$. Again by Lemma 
\ref{new-1017c}, $\{+a\st a\in M'\}\cup\{-a\st a\notin M\}$ is closed
under $aic(P)$. It follows that $\{+a\st a\in M'\}\cup\{-a\st a\notin 
M\}=\{+a\st a\in M\}\cup\{-a\st a\notin M\}$. Thus, $M'=M$ and so, $M$
is a minimal model of $P^M$, that is, an answer set of $P$.
~\hfill$\Box$

We now move on to results concerning upper bounds (membership) and 
derive the main results of this section. 

\begin{lemma}
\label{cl1}
Let $\eta$ be a finite set of normal 
active integrity constraints 
and let $\U$ be a finite set of update actions. 
There is the least set of
update actions $\W$ such that $\U\subseteq \W$ and $\W$ is closed under
$\eta$.  Moreover, this least set $\W$ can be computed in polynomial
time in the size of $\eta$ and $\U$.
\end{lemma}
\textbf{Proof:} We prove the result by demonstrating a bottom-up process
computing $\W$. The process is similar to that applied when computing 
a least model of a Horn program. We start with $\W_0=\U$, Assuming that
$\W_i$ has been computed, we identify in $\eta$ every active integrity
constraint $r$ such that $\nup(r)\subseteq
\lit(\W_i)$, and add the head
of each such rule $r$ to $\W_i$. We call the result $\W_{i+1}$. If
$\W_{i+1}=\W_i$, we stop. It is straightforward to prove that the last
set constructed in the process is closed under $\eta$, contains $\U$,
and is contained in every set that is closed under $\eta$ and contains
$\U$. Moreover, the construction can be implemented to run in polynomial
time. ~\hfill$\Box$

\begin{theorem}
\label{jrc}
Let $\I$ be a database and $\eta$ a set of normal active integrity
constraints. Then checking if there exists a justified repair
(justified weak repair, respectively) for $\<\I,\eta\>$ is an NP-complete
problem.
\end{theorem}
\textbf{Proof:} By Theorem \ref{tn}, it is enough to prove the result
for justified weak repairs.

\smallskip
\noindent
\textsc{(Membership)} The following algorithm decides the problem:
(1) Nondeterministically guess a consistent set of update actions $\E$.
(2) Compute $\nef(\I,\I\circ\E)$.
(3) If $\E\cap\nef(\I,\I\circ\E)\not=\emptyset$
return NO. Otherwise, compute the least set $\W$ of update actions that
is closed under $\eta$ and contains $\nef(\I,\I\circ\E)$.
(4) If $\W=\E\cup
\nef(\I,\I\circ\E)$, then return YES. Otherwise, return NO. From Lemma
\ref{cl1}, it follows that the algorithm runs in polynomial
time. From Theorem \ref{cor-930}, it follows that the algorithm is
correct.

\smallskip
\noindent
\textsc{(Hardness)}
The problem of the existence of an answer set of a simple normal logic
program $P$ is NP-complete. By Theorem \ref{tn} and Theorem
\ref{new-1017b}, $P$ has an answer set if and only if there exists a
justified weak repair for $\<\emptyset,aic(P)\>$. Since $aic(P)$ can
be constructed in polynomial time in the size of $P$, the result
follows.~\hfill $\Box$

\begin{lemma}
\label{cl2}
Let $\eta$ be a finite set of  
active integrity constraints
and let $\U'$ and $\U''$ be sets of update actions.
The problem whether 
there is a set $\U$ of update actions such that $\U$ is closed under
$\eta$ and $\U'\subseteq\U\subset \U''$  is in NP.
\end{lemma}
\textbf{Proof:} Once we nondeterministically guess $\U$, checking all the
required conditions can be implemented in polynomial time. ~\hfill$\Box$

\begin{lemma}
\label{cl3}
Let $\eta$ be a finite set of 
active integrity constraints,
$\I$ a database, and $\E$ be a set of update actions. The problem 
whether there is a set $\E'\subset\E$ of update actions such that $\I\circ\E'\models\eta$
is in NP.
\end{lemma}
\textbf{Proof:} Once we nondeterministically guess $\E$, checking all the 
required conditions can be implemented in polynomial time.
 ~\hfill$\Box$

\begin{theorem}
\label{jwrsp2}
Let $\I$ be a database and $\eta$ a set of active integrity constraints.
The problem of the existence of a justified weak repair for $\<\I,\eta\>$
is a $\Sigma_2^P$-complete problem.
\end{theorem}
\textbf{Proof:}
\textsc{(Membership)} The problem can be decided by a nondeterministic
polyno\-mial-time Turing Machine with an NP-oracle. Indeed, in the first
step, one needs to guess (nondeterministically) a consistent set $\E$
of update 
actions. Setting $\U=\E\cup\nef(\I,\I\circ\E)$, one needs to verify that 
\begin{enumerate}
\item $\E\cap\nef(\I,\I\circ\E)= \emptyset$
\item $\U$ is closed under $\eta$
\item for each $\U'$ such that $\nef(\I,\I\circ\E)\subseteq\U'\subseteq
\U$ and $\U'$ closed under $\eta$, $\U'=\U$ (by Lemma \ref{cl2}, one call
to an NP-oracle suffices).
\end{enumerate}
\textsc{(Hardness)}
The problem of the existence of an answer set of a simple disjunctive
logic program $P$ is $\Sigma_2^P$-complete. By Theorem \ref{new-1017b},
$P$ has an answer set if and only if there exists a justified weak 
repair for $\<\emptyset,aic(P)\>$. Thus, the result follows.
~\hfill $\Box$

\begin{theorem}
\label{jrsp2}

Let $\I$ be a database and $\eta$ a set of active integrity constraints.
The problem of the existence of a justified repair for $\<\I,\eta\>$
is a $\Sigma_2^P$-complete problem.
\end{theorem}
\textbf{Proof:}
\textsc{(Membership)} The problem can be decided by a nondeterministic
polyno\-mial-time Turing Machine with an NP-oracle. Indeed, in the first
step, one needs to guess (nondeterministically) a consistent set $\E$
of update actions. Setting $\U=\E\cup\nef(\I,\I\circ\E)$, one needs to 
verify that 
\begin{enumerate}
\item $\E\cap\nef(\I,\I\circ\E)= \emptyset$
\item $\U$ is closed under $\eta$
\item for each $\U'$ such that $\nef(\I,\I\circ\E)\subseteq\U'\subseteq
\U$ and $\U'$ closed under $\eta$, $\U'=\U$ (by Lemma \ref{cl2}, one 
call to an NP-oracle suffices)
\item for each $\E'$ such that $\E'\subset \E$, $\I\circ \E'\not\models
\eta$ (By Lemma \ref{cl3}, one call to an NP-oracle suffices).
\end{enumerate}
\textsc{(Hardness)} Since for the class of instances $\<\emptyset,aic(P)
\>$ justified weak repairs coincide with justified repairs (Theorem 
\ref{td}), the result follows. 
~\hfill $\Box$

\section{Some implications of the results obtained so far}
\label{summaryRP}

\begin{figure}
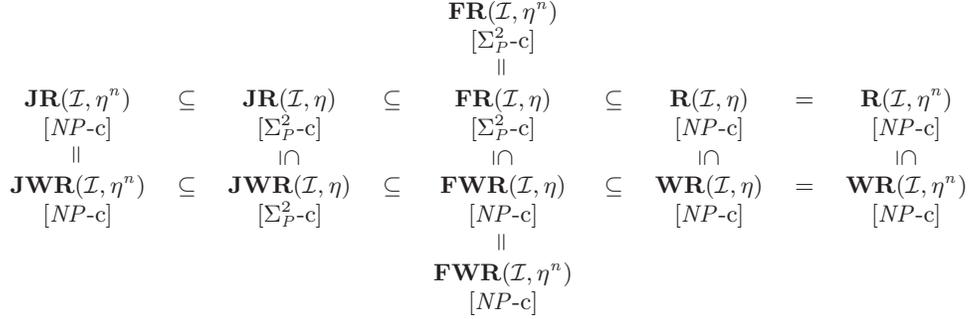

\begin{center}
{\small
\begin{tabular}{ccccccccc}
& &  & & $\bFR(\I,\eta^n)$ & & & &\\ 
& &  & & [$\Sigma_P^2$-c] & & & & \\
& &  & & \veq & & & & \\
$\bJR(\I,\eta^n)$ & $\subseteq$ & $\bJR(\I,\eta)$ & $\subseteq$ & $\bFR(\I,\eta)$ & $\subseteq$ & $\bR(\I,\eta)$ & $=$ & $\bR(\I,\eta^n)$ \\
\mbox{[$NP$-c]} & & \mbox{[$\Sigma_P^2$-c]} & & \mbox{[$\Sigma_P^2$-c]} & & \mbox{[$NP$-c]} & & \mbox{[$NP$-c]}\\ 
\veq & & \vinc & & \vinc & & \vinc & & \vinc\\
$\bJWR(\I,\eta^n)$&$\subseteq$ & $\bJWR(\I,\eta)$ & $\subseteq$ & $\bFWR(\I,\eta)$ & $\subseteq$ & $\bWR(\I,\eta)$ & $=$ & $\bWR(\I,\eta^n)$\\
\mbox{[$NP$-c]}&$ $ & \mbox{[$\Sigma_P^2$-c]} & $ $ & \mbox{[$NP$-c]} & $ $ & \mbox{[$NP$-c]} & $ $ & \mbox{[$NP$-c]}\\
& &  & & \veq & & & & \\
& &  & & $\bFWR(\I,\eta^n)$ & & & & \\
& &  & & \mbox{[$NP$-c]} & & & & 
\end{tabular}
}
\caption{Relationships among classes of repairs} 
\label{fig1}
\end{center}
\end{figure}

We recall that given a database $\I$ and a set $\eta$ of 
aic's, the goal is to replace $\I$ with $\I'$ so that $\I'$ 
satisfies $\eta$. The set of update actions needed to transform $\I$ into
$\I'$ must at least be a repair for $\<\I,\eta\>$ (assuming we insist on
change-minimality, which normally is the case). However, it should 
also obey preferences captured by the heads of constraints in $\eta$. 
Let us denote by $\bR(\I,\eta)$, $\bWR(\I,\eta)$, $\bFR(\I,\eta)$, 
$\bFWR(\I,\eta)$, $\bJR(\I,\eta)$, and $\bJWR(\I,\eta)$ 
the classes of repairs, weak repairs, founded repairs, founded weak 
repairs, justified repairs and justified weak repairs for $\<\I,\eta\>$, 
respectively. Figure \ref{fig1} shows the relationships among these 
classes, with all inclusions being in general proper. Under each class we
also give the complexity of deciding whether a repair from that class exists. 

Thus, given an instance $\<\I,\eta\>$ of the database repair problem, 
one might first attempt to select a repair for $\<\I,\eta\>$ from the
most restricted set of repairs, $\textbf{JR}(\I,\eta^n)$. Not only
these repairs are strongly tied to preferences expressed by $\eta$ ---
the related computational problems are relatively easy. The problem to 
decide whether $\bJR(\I,\eta^n)$ is empty is NP-complete. However, the class 
$\textbf{JR}(\I, \eta^n)$ is narrow and it may be that $\textbf{JR}
(\I,\eta^n)=\emptyset$.
If it is so, the next step might be to try to repair $\I$ by
selecting a repair from $\textbf{JR}(\I,\eta)$. This class of repairs 
for $\<\I,\eta\>$ reflects the preferences captured by $\eta$. Since
it is broader than the previous one, there is a better possibility it
will be non-empty. However, the computational complexity grows --- the 
existence problem for $\textbf{JR}(\I,\eta)$ is $\Sigma_P^2$-complete.
If also $\textbf{JR}(\I,\eta)=\emptyset$, it still may be that founded 
repairs
exist. Moreover, deciding whether a founded repair exists is not harder
than the previous step. Finally, if there are no founded repairs, one
still may consider just a repair. This is not quite satisfactory as it
ignores the preferences encoded by $\eta$ and concentrates only on
the constraint enforcement. However, deciding whether a repair exists is
``only'' NP-complete. Moreover, this class subsumes all other classes of
repairs and offers the best chance of success.

We note that if we fail to find a justified or founded repair in the 
process described above, we may decide that respecting preferences 
encoded in aic's is more important than the 
minimality of change postulate. In such case, rather to proceed to seek
a repair, as discussed above, we also have an option to consider 
justified weak repairs of $\<\I,\eta\>$, where the existence problem 
is $\Sigma_2^P$-complete and, then founded weak repairs for $\<\I,\eta\>$,
where the existence problem is NP-complete.

Finally, we point out that when we choose a smaller class $\A$ of repairs 
(e.g., $\textbf{JR}(\I,\eta^n)$) instead of a more general one $\A'$
(e.g., $\textbf{JR}(\I,\eta)$) we observe an important (and desirable)
effect on \emph{consistent query answering}. In consistent query answering
(\emph{conservative reasoning}) an atom $a$ is \emph{true} if it belongs to every
repaired database, \emph{false} if it does not belong to any repaired 
database, and \emph{unknown} otherwise, that is, when it belongs to a
\emph{proper} subset of the set of repaired databases. It is clear that
if $\A\subseteq\A'$, the set of the true atoms under $\A'$ is a subset 
of the set of the true atoms obtained by applying $\A$, and the same holds
for atoms that are false. In other words, the stronger the semantics, the 
larger the set of atoms that receive a definite truth value (are true or
false). 

\section{Revision Programming --- an Overview}

We review the basic terminology of revision programming,
and recall the two semantics introduced by Marek, Truszczy\'{n}ski and 
Pivkina \citeyear{mt94b,mt94c,piv01}: 
the semantics
of supported revisions, and the semantics of justified weak revisions
(originally referred to as justified revisions and 
renamed here for consistency with the general naming schema we use).

\noindent
\textbf{Revision literals.}
A \emph{revision literal} is an expression $\rin(a)$ or $\rout(a)$,
where $a\in\At$. Revision literals $\rin(a)$ and $\rout(a)$ are
\emph{duals} of each other. If $\alpha$ is a revision literal, we denote
its dual by $\alpha^D$. We extend this notation to sets of revision 
literals. We say that a set of revision literals is \emph{consistent}
if it does not contain a pair of dual literals.
Revision literals represent elementary updates one can apply to a
database. We define the result of applying a \emph{consistent} set 
$\U$ of revision literals to a database $\I$ as follows:
\[
\I\oplus\U = (\I\cup\{a\st \rin(a)\in\U\})\setminus\{a\st\rout(a)\in
\U\}.
\]

\noindent
\textbf{Revision rules, normal rules and constraints.}
A \emph{revision rule} is an expression of the form
\begin{equation}
\label{RevRule}
r = \alpha_1|\ldots|\alpha_k\leftarrow\beta_1,\ldots,\beta_m,
\end{equation}
where $k,m\geq 0$, $k+m\geq 1$, and $\alpha_i$ and $\beta_j$ are 
revision literals. The set $\{\alpha_1,\ldots,\alpha_k\}$ is the 
\emph{head} of the rule
(\ref{RevRule}); we denote it by $\hd(r)$. Similarly, the set $\{
\beta_1,\ldots,\beta_m\}$ is the \emph{body} of  the rule
(\ref{RevRule}); we denote it by $\bd(r)$. A revision rule is 
\emph{normal} if $|\hd(r)|\leq 1$. As in the case of active integrity
constraints, we denote the empty head as $\bot$. We call rules with the empty 
head \emph{constraints}.  
If $|\bd(r)|= 0$ we omit the implication symbol. Examples of revision
rules are: (1) $\rin(a)| \rout(b) \leftarrow \rin(c)$, (2) $\rin(a)|\rin(c)$,
(3) $\rin(a)\leftarrow \rout(b)$, and (4) $\bot \leftarrow \rin(a),\rout(b)$.
The second rule is an example of a rule with the empty body, the third one 
is an example of a normal rule and the last one is an example of a constraint.
The informal reading of a revision rule, say the first rule given above, 
$\rin(a)| \rout(b) \leftarrow \rin(c)$, is: \emph{insert $a$ or delete $b$, 
if $c$ is present}.

\noindent
\textbf{Revision programs.}
A \emph{revision program} 
is a collection of revision rules. A revision program is \emph{normal}
if all its rules are normal. 

\noindent
\textbf{Entailment (satisfaction).}
A database $\I$ \emph{satisfies} a revision literal $\rin(a)$ ($\rout
(b)$, respectively), if $a\in\I$ ($b\notin\I$, respectively). A 
database $\I$ \emph{satisfies} a revision rule (\ref{RevRule}) if it
satisfies at least one literal $\alpha_i$, $1\leq i\leq k$, whenever
it satisfies every literal $\beta_j$, $1\leq j\leq m$.
Finally, a database $\I$ satisfies a revision program $P$, if $\I$ 
satisfies every rule in $P$. We use the symbol $\models$ to denote the 
satisfaction relation.

For revision literals $\alpha=\rin(a)$ and $\beta=\rout(b)$, we set
$\lit(\alpha)=a$ and $\lit(\beta)=\nt b$. We extend this
notation to sets of revision literals. We note that every database
interprets revision literals and the corresponding propositional literals
in the same way.
That is, for every database $\I$ and for every set of revision literals $L$,
$\I\models L$ if and only if $\I\models\lit(L)$. 

It follows that a revision rule (\ref{RevRule}) specifies an integrity
constraint equivalent to the propositional formula:
$\lit(\beta_1),\ldots,\lit(\beta_m)\supset\lit(\alpha_1),\ldots,
\lit(\alpha_k)$.
However, a revision rule is not only an integrity
constraint. Through its syntax, it also encodes a preference on how to
``fix'' a database, when it violates the constraint. Not satisfying a revision rule $r$
means satisfying all revision literals in the body of $r$ \emph{and}
not satisfying any of the revision literals in the head of $r$. Thus,
enforcing the constraint means constructing a database that (1) does not
satisfy some revision literal in the body of $r$, \emph{or} (2)
satisfies at least one revision literal in the head of $r$. The
underlying idea of revision programming is to prefer those revisions
that result in databases with the property (2).

As an example, let us consider the revision rule 
$r=\rin(a)\leftarrow \rout(b)$, and the empty database $\I$. 
Clearly, $\I$ does not satisfy $r$. Although $\I$ can be fixed either 
by inserting $a$, 
so that $\hd(r)$ becomes \emph{true}, or by inserting $b$, so that 
$\bd(r)$ becomes \emph{false}, the syntax of $r$ makes the former
preferred. 

Normal revision programs were introduced and studied by Marek and 
Truszczy\'{n}\-ski \citeyear{mt94b,mt94c}, who proposed the syntax and
the semantics of supported and justified weak revisions.
The formalism was extended by Pivkina 
\citeyear{piv01} to allow disjunctions of revision literals in the heads 
of rules, and the semantics of justified weak revisions was generalized 
to that case. We will now recall these definitions.

First, we define the notion of the \emph{inertia set}. Let $\I$ and 
$\R$ be databases. We define the \emph{inertia set} wrt 
$\I$ and $\R$, denoted $I(\I,\R)$, by setting
\[
I(\I,\R)=\{\rin(a)\st a\in\I\cap\R\} \cup \{\rout(a)\st
a\notin\I\cup\R\}.
\]
In other words, $I(\I,\R)$ is the set of all \emph{no-effect} revision
literals for $\I$ and $\R$, that is, revision literals that have no 
effect when revising $\I$ into $\R$.

Now, let $P$ be a \emph{normal} revision program and $\R$ be a database. 
By $P_\R$ we denote the program obtained from $P$ by removing each rule
$r\in P$ such that $\R\not\models\bd(r)$. 

\begin{definition}[\textsc{Supported updates and supported revisions}] Let
$P$ be a normal revision program and $\I$ a database. A set $\U$ of 
revision literals is a \emph{supported update} of $\I$ wrt 
$P$ if $\U$ is consistent and $\U=\hd(P_{\I\oplus\U})$. A set $\E$ is a 
\emph{supported revision} of $\I$ wrt $P$ if $\E=\U
\setminus I(\I,\I\oplus\U)$,
where $\U$ is a supported update.
~\hfill$\Box$
\end{definition}
Intuitively, a consistent set $\U$ of revision literals is a supported 
update if it is precisely the set of literals ``supported'' by $P$ and
the database resulting from updating $\I$ with $\U$. Eliminating from
a supported revision all no-effect literals yields a supported revision.

While not evident explicitly from the definition, supported updates
and revisions guarantee constraint enforcement, as proved by Marek
and Truszczy\'nski \citeyear{mt94c}.

\begin{proposition}
\label{prop14}
Let $P$ be a normal revision program and $\I$ a database. If $\E$ is a 
supported revision of $P$, then $\I\oplus\E\models P$.~\hfill $\Box$
\end{proposition}

Supported updates do not take into account the inertia set. Supported 
revisions do, but only superficially: simply removing no-effect literals 
from the corresponding supported update. It is then not surprising that
supported updates and revisions may be self-grounded and non-minimal,
as we show in the following example.

\begin{example}
Let $P$ be a revision program containing the rules 
$\{\rin(a)\leftarrow \rin(b),\  \rin(b)\leftarrow \rin(a),$  
$\rin(c)\leftarrow \rout(d)\}$, 
and let $\I$ the empty database. 
$\I$ does not satisfy $P$ as it violates the rule $\rin(c)\leftarrow \rout(d)$. 
One can check that set $\U=\{\rin(a),\rin(b),\rin(c)\}$
modeling the insertions of $a$, $b$
and $c$, is a supported update and a supported revision.
However it is not minimal as its subset $\{\rin(c)\}$ is sufficient
to guarantee the satisfaction of $P$.~\hfill$\Box$
\end{example}

The problem in the previous example is self-groundedness or the circularity 
of support between 
$\rin(a)$ and $\rin(b)$. Each of them supports the other one but 
the set containing both is superfluous.
To address the problem, Marek and Truszczy\'{n}ski \citeyear{mt94b,mt94c} 
proposed for normal revision programs
the semantics of justified weak revisions, later extended to the 
disjunctive case by Pivkina \citeyear{piv01}. The idea was to ``ground'' 
justified weak revisions in
the program and the inertia set by means of a \emph{minimal closure}.

\begin{definition}[\textsc{Minimal closed sets of revision literals}]
A set $\U$ of revision literals is \emph{closed} under a revision 
program $P$ (not necessarily normal) if for every rule $r\in P$, 
whenever $\bd(r)\subseteq \U$, then $\hd(r)\cap \U\not=\emptyset$.
If $\U$ is closed under $P$ and for every set $\U'\subseteq\U$ closed
under $P$, we have $\U'=\U$, then $\U$ is a \emph{minimal closed}
set for $P$.
~\hfill$\Box$
\end{definition}

With this definition in hand, we can define the concepts
of justified updates and justified weak revisions.

\begin{definition}[\textsc{Justified updates and justified weak
revisions}]
\label{WeaklyJustRev}
Let $P$ be a revision program and let $\I$ be a database. A consistent
set $\U$ of revision literals is a \emph{$P$-justified update} for $\I$ if
it is a minimal set closed under $P\cup I(\I,\I\oplus\U)$.

If $\U$ is a $P$-justified update for $\I$, then $\U\setminus
I(\I,\I\oplus\U)$ is a \emph{$P$-justified weak revision} for $\I$.
~\hfill$\Box$
\end{definition}
We note that $P\cup I(\I,\I\oplus\U)$ is well defined as revision 
literals (and so, in particular, the revision literals in $I(\I,\I\oplus\U)$)
are special revision rules (normal and with empty bodies).

The inertia set plays an essential role in the definition,
as it is used directly in the definition of a $P$-justified update.
Again, it is not self-evident from the definition that justified
updates and justified weak revisions, when applied to an initial 
database yield a database satisfying the program. However, the definition 
does indeed imply so \citep{mt94c,piv01}.

\begin{proposition}
\label{prop15}
Let $P$ be a revision program and $\I$ a database. If $\U$ is a
justified update or justified weak 
revision of $P$, then $\I\oplus\U\models P$.~\hfill$\Box$
\end{proposition}

We point out that the original term for the \emph{justified weak revisions}
was \emph{justified revisions} \citep{mt94c}. We changed the name for consistency
with the naming schema we used for active integrity constraints.

\section{A Family of Declarative Semantics for Revision Programming}

The two semantics in the previous section were defined based on how revisions 
are ``grounded'' in a program, an initial database, and the inertia set. The 
fundamental postulates of constraint enforcement and minimality of change 
played no explicit role in those considerations. The first one is no problem 
as it is a side effect of each of the two types of groundedness considered 
(cf. Propositions \ref{prop14} and \ref{prop15}). The second one does not 
hold for supported revisions. And while Marek and Truszczy\'nski 
\citeyear{mt94c} proved that justified weak revisions are change-minimal in 
the case of \emph{normal} revision programs, it is not so in the general case. 

\begin{example}
\label{ex132}
Let $P$ be a revision program consisting of the rules
$\rin(a)|\rout(b)$, $\rout(a)|\rin(b)$, and let $\I$ be the empty database.  
It is easy to verify that set $\{\rin(a),\rin(b)\}$ is a justified
weak revision. However, it is not minimal as $\I$ is already consistent and 
no update is needed (or, in other words, the empty update fixes the 
consistency). 
~\hfill$\Box$
\end{example}

We will now develop a range of semantics for revision programs by taking 
the postulates of constraint enforcement and minimality of change explicitly 
into consideration. 

\begin{definition}[\textsc{Weak Revisions and Revisions}]
A consistent set $\U$ of revision literals is a \emph{weak revision} of $\I$
wrt a revision program $P$ if 
(1) $\U\cap I(\I,\I\oplus\U)=\emptyset$ (relevance --- all revision 
literals in $\U$ actually change $\I$ or, in other words, none of them
is a no-effect literal wrt $\I$ and $\I\oplus\U$);
and
(2) $\I\oplus\U\models P$ (constraint enforcement).
Further, $\U$ is a \emph{revision} of $\I$
with respect to a revision program $P$ if it is a weak revision and for every
$\U' \subseteq\U$, $\I\oplus\U'\models P$ implies that $\U'=\U$ 
(minimality of change).~\hfill$\Box$
\end{definition}

\begin{example}
\label{exn}
Let $P$ be the program consisting of the two rules from Example
\ref{ex132} and the rule $\rin(c)\leftarrow \rout(d)$. As before, let $\I=
\emptyset$. There are several weak revisions of $\I$ with respect to $P$, 
for instance, $U_1=\{\rin(d)\}$, $\U_2=\{\rin(d),\rin(a),\rin(b)\}$, 
$\U_3=\{\rin(c)\}$, and $\U_4=\{\rin(c),\rin(a),\rin(b)\}$. The weak
revisions $\U_1$ and $\U_3$ are minimal and so, they are revisions. 
~\hfill$\Box$
\end{example}

(Weak) revisions do not reflect the preferences on how to revise
a database encoded in the syntax of revision rules. 
Justified weak revisions and supported revisions, which we discussed in
the previous section, do.

\smallskip
\noindent
\textit{Example \ref{exn} (continued)}\\
Both the semantics of supported revisions and justified weak revisions
exclude the weak revisions $\U_1=\{\rin(d)\}$ and $\U_2=\{\rin(d),\rin(a),
\rin(b)\}$, in favor of $\U_3=\{\rin(c)\}$ and $\U_4=\{\rin(c),\rin(a),
\rin(b)\}$ ($\U_3$ and $\U_4$ indeed are supported and justified weak revisions), thus \emph{preferring} to satisfy the head of the rule $\rin(c)
\leftarrow \rout(d)$ rather than to violate its the body. Indeed, one can
check that $\U_3$ and $\U_4$ are indeed both supported and justified weak
revisions, while $\U_1$ and $\U_2$ are neither.~\hfill$\Box$
\smallskip

We will now introduce several additional semantics that aim to capture 
this preference.
First, we define a new semantics for revision programs by strengthening
the semantics of justified weak revisions. We do so simply by imposing 
change-minimality explicitly. 

\begin{definition}[\textsc{Justified Revisions}]
\label{JustRev}
Let $P$ be a revision program and let $\I$ be a database.
A $P$-justified weak revision $\E$ for $\I$ is a 
\emph{$P$-justified revision} for $\I$ if $\E$ is a revision of $\I$ 
wrt $P$ (that is, for every set $\E' \subseteq \E$ such that
$\I\oplus\E'\models P$, $\E'=\E$).
~\hfill$\Box$
\end{definition}

\noindent
\smallskip
\textit{Example \ref{exn} (continued)}\\
Let us consider again Example \ref{exn}. The set $\U_3$ is a $P$-justified
revision for $\I$, while $\U_4$ is not, reflecting the fact that we require
that $P$-justified revisions be revisions (that is, satisfy change minimality).
~\hfill $\Box$

\smallskip
Justified revisions have several useful properties. They are 
change-minimal and are grounded in the program \emph{and} the inertia
set. However, as stable models of logic programs, to which they are closely
related, in some settings they may be too restrictive.

\begin{example}
\label{exf}
Let $P=\{\rin(a) \leftarrow\ \rin(a),\ \rin(a)\leftarrow\ \rout(a)\}$
and let $\I=\emptyset$. Clearly, $\I$ is inconsistent with respect to
$P$. The set $\U=\{\rin(a)\}$ is a revision of $\I$ and one might argue
that $P$ provides it a justification: the two rules together ``force'' $a$
into $\I$, as in any particular situation one of them applies and provides
a justification for $\rin(a)$. This type of an argument is known as 
``reasoning by cases.'' However, one can check that $\U$ is not a 
$P$-justified revision of $\I$ and not a $P$-justified weak revision, either. 
Thus, justified (weak) revisions in general exclude such reasonings as valid.
~\hfill $\Box$
\end{example}

To provide a semantics capturing such justifications, we introduce now 
the concept of foundedness and the semantics of founded (weak) revisions.
We follow closely intuitions behind founded (weak) repairs.

\begin{definition}[\textsc{Founded (weak) revisions}]
\label{def::actconstsem3rev}
Let $\I$ be a database, $P$ a revision program and,
and $\E$ a consistent set of revision literals.
\begin{enumerate}
\item
A revision literal $\alpha$ is $P$-\emph{founded} wrt $\I$
and $\E$ if there is $r\in P$ such that $\alpha\in\hd(r)$, $\I\oplus\E
\models\bd(r)$, and $\I\oplus\E\models\beta^D$, for every $\beta\in
\hd(r)\setminus\{\alpha\}$.
\item
The set $\E$ is $P$-\emph{founded} wrt $\I$
if every element of $\E$ is $P$-founded wrt $\I$
and $\E$.
\item
$\E$ is a $P$-\emph{founded (weak) revision} for $\I$ if $\E$ is a
(weak) revision of $\I$ wrt $P$ and $\E$ is
$P$-\emph{founded} wrt $\I$.~\hfill $\Box$
\end{enumerate}
\end{definition}

It is clear from the definition that $P$-foundedness of a revision literal 
$\alpha$ with respect to a consistent set of revision literals $\E$ can be 
established by considering rules in $P$ independently of each other, which
supports reasoning by cases such as the one used in Example \ref{exf} (in 
this specific case, $\rin(a)$ is founded either because of the first rule
or becaue of the second rule). Indeed,
one can verify that the revision $\U$ in Example \ref{exf} is founded. 

We note that condition (3) of the definition guarantees that founded (weak)
revisions enforce constraints of the revision program. Next, directly from 
the definition, it follows that founded weak revisions are weak revisions. 
Similarly, founded revisions are revisions and so, they are change-minimal. 
Furthermore, founded revisions are founded weak revisions. However, there 
are (weak) revisions that are not founded, and founded weak revisions are not 
necessarily founded revisions, that is, are not change-minimal. The 
latter observation shows that foundedness is too weak a condition to 
guarantee change-minimality. 

\begin{example}
\label{ex-925c}
Let $P$ be the revision program containing the rules 
$\{\rin(b) \leftarrow \rin(a),\ \rin(a) \leftarrow \rin(b),
\rin(c) \leftarrow \rout(d)\}$
and $\I$ the empty database.
The set $\{\rin(d)\}$ is a revision of $\I$ wrt $P$.
Therefore it is a weak revision of $\I$ wrt $P$. However,
it is not a $P$-founded weak revision for $\I$. Therefore, it is not a
$P$-founded revision for $\I$, either. The set $\{\rin(c),\rin(a),
\rin(b)\}$ is a $P$-founded weak revision for $\I$ but not a $P$-founded
revision for $\I$. Indeed, $\{\rin(c)\}$ is also a revision of $\I$
wrt $P$.~\hfill $\Box$
\end{example}

In the case of normal revision programs, founded weak revisions coincide
with supported revisions.

\begin{theorem}
\label{fwrsr}
Let $P$ be a normal revision program and $\I$ a database. A set $\E$ of
revision literals is a $P$-founded weak revision of $\I$ if and only if
$\E$ is a $P$-supported revision of $\I$.  ~\hfill $\Box$
\end{theorem}
\textbf{Proof:}$(\Rightarrow)$
Let $\E$ be a $P$-founded weak revision of $\I$ and let
$\U=\E\cup (I(\I,\I\oplus\E) \cap \hd(P_{\I\oplus\E}))$.
As $\E$ is a weak revision of $\I$ with respect to $P$, $\E\cap
I(\I,\I\oplus\E)=\emptyset$. Therefore, $\E=\U\setminus I(\I,\I
\oplus\E)$ and $\I\oplus\E=\I\oplus\U$. It follows that $\E=\U\setminus
I(\I,\I\oplus\U)$ and so, it will suffice to prove that $\U$ is a 
supported update of $\I$ with respect to $P$. 

To this end, we first note that $\U$ is consistent. Indeed:
\begin{enumerate}
	\item $\E$ is consistent (it is a weak revision);
	\item $I(\I,\I\oplus\E)$ is consistent;
	\item If $\alpha\in\E$ then the literal $\alpha^D\notin
              I(\I,\I\oplus\E)$.
\end{enumerate}
Next, we prove that $\U=\hd(P_{\I\oplus\U})$. Let $\alpha\in\U$. We have
two cases: either $\alpha \in I(\I,\I\oplus\E) \cap \hd(P_{\I\oplus\E})$
or $\alpha\in\E$. The first case trivially verifies the assertion.
In the second case, as $\E$ is a $P$-founded weak revision of $\I$,
there exists $r\in P$ such that $\alpha=\hd(r)$ and $\I\oplus\E 
\models \bd(r)$ (cf. Definition 5). Thus, $r\in P_{\I\oplus\E}$ and
$\alpha\in\hd(P_{\I\oplus\E})$. As $\I\oplus\E=\I\oplus\U$ we have 
$\alpha\in\hd(P_{\I\oplus\U})$.

Conversely, let $\alpha\in\hd(P_{\I\oplus\U})$. We have two cases: $\alpha
\in I(\I,\I\oplus\E)$, and $\alpha \not\in I(\I,\I\oplus\E)$. In the first
case, $\alpha\in\U$ (by the definition of $\U$). In the second case, we 
reason as follows. Since $\alpha\in\hd(P_{\I\oplus\U})$, there exists 
$r\in P$ such that $\alpha=\hd(r)$ and $\I\oplus\U \models\bd(r)$. Thus,
$\I\oplus\E\models\bd(r)$.
As $\E$ is a weak revision, $\I\oplus\E\models r$. Consequently, 
$\I\oplus\E\models\alpha$. Since $\alpha\notin I(\I,\I\oplus\E)$, 
$\I\not\models\alpha$ and so, $\alpha\in\E$. Thus, $\alpha \in \U$.

\smallskip
\noindent
$(\Leftarrow)$
Let $\E$ be a $P$-supported revision of $\I$. It follows that $\E=
\U\setminus I(\I,\I\oplus\U)$, where $\U$ is a $P$-supported update of
$\I$ wrt $P$. It follows that $\I\oplus\E=\I\oplus\U$. Consequently,
$\E\cap I(\I,\I\oplus\E)=\emptyset$ and, by Proposition \ref{prop14}, 
$\I\oplus \E\models P$. Since $\E\subseteq\U$, $\E$ is consistent and
so, $\E$ is a weak revision of $P$. 

Let $\alpha\in\E$. As $\E \subseteq\U$, there exists $r\in P$ such that
$\alpha=\hd(r)$ and $\I\oplus\U\models\bd(r)$. Thus, $\I\oplus\E\models
\bd(r)$, too. Consequently, $\alpha$ is $P$-founded wrt $\I$ and $\E$.
It follows that $\E$ is a $P$-founded weak revision of $\I$.
~\hfill$\Box$

At an intuitive level, we already argued earlier that foundedness is less 
restrictive than the condition defining justified updates, which is behind 
justified (weak) revisions. We will now make this intuition formal.

\begin{theorem}
\label{theorem:j(w)r-f(w)r}
Let $P$ be a revision program and let $\I$ be a database. If a set $\E$ of
revision literals is 
a $P$-justified (weak) revision of $\I$,
then it is a $P$-founded
(weak) revision of $\I$. 
\end{theorem}
\textbf{Proof:} Let $\E$ be a $P$-justified weak revision of $\I$.
By Proposition \ref{prop15}, $\I\oplus\E\models P$. Moreover, there
is a $P$-justified update $\U$ for $\I$ such that $\E=\U\setminus
I(\I,\I\oplus\U)$. It follows that $\I\oplus\U=\I\oplus\E$ and 
$\E\cap I(\I,\I\oplus\E)=\emptyset$. Since $\U$ s consistent (by the
definition), $\E$ is consistent and so, $\E$ is a weak revision of
$\I$ with respect to $P$. 

To show that $\E$ is a $P$-founded weak revision of $\I$, we need to
prove that $\E$ is $P$-founded wrt $\I$. Let $\alpha\in \E$. We recall
that by the definition, $\U$ is a minimal set closed under $P\cup 
I(\I,\I\oplus\U)$. As $\U$ is minimal, $\U'=\U\setminus\{\alpha\}$ is 
not closed under $P\cup I(\I,\I\oplus\U)$. As $\alpha\not\in I(\I,\I
\oplus\U)$ there is a revision rule $r\in P$ such that $\bd(r)\subseteq
\U'$ and $\hd(r)\cap\U'=\emptyset$. Since $\U'\subseteq\U$, $\bd(r)
\subseteq \U$. It follows that $\I\oplus \U\models\bd(r)$ and so, $\I
\oplus\E\models bd(r)$. 

We recall that $\U$ is closed under $P$. Thus, $\hd(r)\cap\U=\{\alpha\}$.
Let $\beta\in\hd(r)\setminus\{\alpha\}$. It follows that $\beta\not
\in\U$ and so, $\beta\not\in \E$ and $\beta\not\in I(\I,\I\oplus\U)$.
If $\I\models\beta$, then $\beta\not\in I(\I,\I\oplus\U)$ implies that 
$\I\oplus\U\not\models\beta$. If $\I\not\models \beta$, then $\beta\not
\in\E$ implies $\I\oplus\E\not\models\beta$. In each case $\I\oplus\E
\models \beta^D$. It follows that $\alpha$ is $P$-founded wrt $\I$.
Thus, $\E$ is $P$-founded wrt $\I$ and so, it is a $P$-founded weak
 revision of $\I$

Next, let us assume that $\E$ is a $P$-justified revision of $\I$.
Then, $\E$ is a $P$-justified weak revision of $\I$ and so, a
$P$-founded weak revision of $\I$ (by the argument above). In 
particular, it is $P$-founded wrt $\I$. Moreover, since $\E$ is a 
$P$-justified revision of $\I$, it is a revision of $\I$ wrt
$P$. Therefore, $\E$ it is a $P$-founded revision of $\I$ wrt $P$. 
~\hfill$\Box$

The converse implications do not hold in general (cf. Example \ref{exf}).

As in the case of active integrity constraints, revision rules can be
\emph{normalized}. Namely, for a revision rule
$
r=\alpha_1|\ldots|\alpha_k\leftarrow \phi
$ 
by $r^n$ we denote the set of \emph{normal} revision rules as follows:
$r^n=\{r\}$, if $k\leq 1$ or, if $k\geq 2$, $r^n=\{r_1,\ldots,r_k\}$, 
where $r_i = \alpha_i \leftarrow \alpha_1^D,\ldots,\alpha_{i-1}^D,
\alpha_{i+1}^D,\ldots,\alpha_k^D,\phi$. For a revision program $P$,
we define $P^n=\bigcup_{r\in P} r^n$. One can prove the following result
(we omit the details as they are quite similar to those we presented above).

\begin{theorem}
\label{normRP}
Let $P$ be a revision program and let $\I$ be a database. A set $\E$ of
revision literals is a (weak) revision of $\I$ with respect to $P^n$
($P^n$-founded (weak) revision of $\I$, respectively) if and only if 
it is a (weak) revision of $\I$ with respect to $P$ ($P$-founded (weak)
revision of $\I$, respectively). Moreover, if $\E$ is a $P^n$-justified 
(weak) revision of $\I$, then it is a $P$-justified (weak) revision of 
$\I$. 
\end{theorem}  

To summarize our discussion so far, revision programs can be assigned the
semantics of (weak) revisions, justified (weak) revisions and
founded (weak) revisions. Thanks to Theorem \ref{normRP}, we can also
assign to a revision program $P$ the semantics of $P^n$-justified revisions. 
Let us denote the classes of the corresponding types of revisions by 
$\bRev(\I,P)$, $\bWRev(\I,P)$,  
$\bJRev(\I,P)$, $\bJWRev(\I,P)$,
$\bFRev(\I,P)$ and $\bFWRev(\I,P)$.
The relationships between the semantics we discussed above are demonstrated 
in Figure \ref{fig1a}. One can show that none of the containment relations
can be replaced with the equality.
\begin{figure}
\centerline{\tiny{
\begin{tabular}{ccccccccc}
& & & & $\bFRev(\I,P^n)$ & & & $ $& \\
& & & & \veq & & & & \\
$\bJRev(\I,P^n)$ & $\subseteq$& $\bJRev(\I,P)$ & $\subseteq$ & $\bFRev(\I,P)$ & $\subseteq$ & $\bRev(\I,P)$& $=$& $\bRev(\I,P^n)$\\
$\veq$ & & \vinc & & \vinc & & \vinc& &\vinc \\
$\bJWRev(\I,P^n)$ & $\subseteq$& $\bJWRev(\I,P)$ & $\subseteq$ & $\bFWRev(\I,P)$ & $\subseteq$ & $\bWRev(\I,P)$& $=$& $\bWRev(\I,P^n)$\\
& & & & \veq & & & & \\
& & & & $\bFWRev(\I,P^n)$ & & & & 
\end{tabular}
}}
\caption{The containment relations for the semantics of revision programs} 
\label{fig1a}
\end{figure}

The similarities revision programs show to sets of active integrity 
constraints are striking. In the next section, we will now establish 
the precise connection.

\section{Connections between Revision Programs and Active Integrity
Constraints}
\label{connections}
To relate revision programs and active integrity constraints, we first
note that we can restrict the syntax of revision programs without 
affecting their expressivity. 

A \emph{proper revision rule} is a revision rule that satisfies the
following condition: any literal in the head is not the
dual of any literal in the body.

Let $P$ be a revision program and let $r_1$ and $r_2$
be revision rules
\[
\alpha|\alpha_1|\ldots|\alpha_k \leftarrow
\alpha^D,\beta_1,\ldots,\beta_m
\]
and
\[
\alpha_1|\ldots|\alpha_k \leftarrow 
\alpha^D,\beta_1,\ldots,\beta_m,
\]
respectively (that is, $r_2$ differs from $r_1$ in that it drops $\alpha$ 
from the head).

\begin{lemma}
\label{properization}
Let $\I$ be a database. Under the notation introduced above, a set of 
revision literals $\U$ is a (weak) revision of $\I$ with respect to $P\cup
\{r_1\}$ ($P\cup\{r_1\}$-founded (weak) revision, $P \cup\{r_1\}
$-justified (weak) revision of $\I$, respectively) if and only if $\U$
is a (weak) revision of $\I$ with respect to $P\cup\{r_2\}$ ($P\cup
\{r_2\}$-founded (weak) revision, $P \cup\{r_2\}$-justified (weak)
revision of $\I$, respectively).
\end{lemma}
\textbf{Proof:} 
The claim is evident for the case of weak revisions and
revisions. The case of justified (weak) revisions follows from the 
observation that a consistent set $\U$ of revision literals is a closed
set for $P\cup\{r_1\}\cup I(\I,\I\oplus\U)$ if and only if $\U$ is a 
closed set for $P\cup\{r_2\}\cup I(\I,\I\oplus\U)$.

For the case of founded (weak) revisions, it is enough to prove that
a set $\U$ of revision literals is $P\cup\{r_1\}$-founded wrt $\I$
if and only if $\U$ is $P\cup\{r_2\}$-founded wrt $\I$. When proceeding
in either direction, we have that $\U$ is consistent. 

Let $\beta\in\U$ be $P\cup\{r_1\}$-founded wrt $\I$ and $\U$,
and let $r\in P\cup\{r_1\}$ be the rule providing support to $\beta$.
If $r\not=r_1$, $r\in P$ and so, $\beta$ is $P\cup\{r_2\}$-founded wrt 
$\I$ and $\U$. Thus, let us assume that $r=r_1$. If $\beta=\alpha$, 
then $\alpha\in\U$ and, consequently, $\I\oplus\U\models\alpha$.
Since $\I\oplus\U\models\bd(r_1)$, $\I\oplus\U\models\alpha^D$, a
contradiction. Thus, $\beta\not=\alpha$. It is easy to see that in such
case, $r_2$ supports $\beta$ (given $\U$). Thus, $\beta$ is $P\cup
\{r_2\}$-founded wrt $\I$ in this case, too. It follows that $\U$ is 
$P\cup\{r_2\}$-founded wrt $\I$.

Conversely, let $\beta\in\U$ be $P\cup\{r_2\}$-founded wrt $\I$ and 
$\U$, and let $r\in P\cup\{r_2\}$ be the rule providing support to 
$\beta$. As before, if $r\not=r_2$, the claim follows. If $r=r_2$,
then $\beta\not=\alpha$. Since $r_2$ supports $\beta$, one can check
that $r_1$ supports, $\beta$, too. Thus, $\beta$ is 
$P\cup\{r_1\}$-founded wrt $\I$ and $\U$. Consequently, $\U$ is
$P\cup\{r_1\}$-founded wrt $\I$
~\hfill$\Box$

Lemma \ref{properization} shows that the literals in the head of
a revision rule which are dual of  literals in the body are useless
and can be dropped. In other words, there is no loss of generality in 
considering just proper revision programs.
\begin{example}
Let $P$ be the revision program containing the rules 
$\{\rin(b)|\rout(a) \leftarrow \rin(a),\rout(d)|$ 
$\rin(c) \leftarrow \rout(c)\}$.
Its properized version is 
$\{\rin(b) \leftarrow \rin(a),$ 
$\rout(d) \leftarrow \rout(c)\}$.
~\hfill $\Box$
\end{example}

\begin{theorem}
\label{properization:thm}
Let $P$ be a revision program. There is a proper revision program $P'$
such that for every database $\I$, (weak) revisions of $\I$ with respect
to $P$ ($P$-founded (weak) revisions, $P$-justified (weak) revisions of 
$\I$, respectively) coincide with (weak) revisions of $\I$ with
respect to $P'$  ($P'$-founded (weak) revisions, $P'$-justified (weak)
revisions of $\I$, respectively).
\end{theorem}
\textbf{Proof:} Lemma \ref{properization} implies that the program $P'$
obtained from $P$ by repeated application of the process described above 
(replacement of rules of the form $r_1$ with the corresponding rules of 
the form $r_2$) has the required property.
~\hfill$\Box$

\noindent
We denote the ``properized'' version of a revision program $P$ as 
$prop(P)$.

\noindent
We extend to revision literals the operator $ua(\cdot)$ defined for 
propositional literals. If $\alpha=\rin(a)$, we define $\ua(\alpha)=+a$. 
If $\alpha=\rout(a)$, we define $\ua(\alpha)=-a$.

\label{equivalence}
\begin{definition}
Given a proper revision rule $r$ of the form
$$
\alpha_1|\ldots|\alpha_k \leftarrow \beta_1,\ldots\beta_m
$$
we denote by $AIC(r)$ the active integrity constraint
$$
\lit(\beta_1),\ldots,\lit(\beta_m),{\lit(\alpha_1)}^D, \ldots,
{\lit(\alpha_k)}^D \supset\act(\alpha_1)|\ldots|\act(\alpha_k).
$$

\vspace*{-0.25in}
~\hfill $\Box$
\end{definition}

For example, given the proper revision rule $r\ :\ \rin(a)\leftarrow 
\rout(b)$, the corresponding active integrity constraint $AIC(r)$ is of 
the form $\nt b,\nt a \supset +a$. We note that if $r$ is a constraint 
($k=0$), $AIC(r)$ is simply an integrity constraint. The
operator $AIC(\cdot)$ is extended to proper revision programs in the 
standard way. It is easy to show that for each database $\I$,
$\I\models P$ if and only if $\I\models AIC(P)$.
The following lemma establishes a direct connection
between the concepts of closure under active integrity constraints and
revision programs.

\begin{lemma}
\label{lemma:closeness}
Let $r$ be a proper revision rule. 
A set $\E$ of revision literals is closed under $P$ if and only if
$\ua(\E)$ is closed under $AIC(r)$.
\end{lemma}
\textbf{Proof:}
First, we observe that as $r$ is proper, $\nup(AIC(r))=lit(\bd(r))$. 
Moreover $\hd(AIC(r))=\ua(\hd(r))$.
We know that $\E$ is closed under $r$ if and only if
$\bd(r)\not\subseteq\E$ or $\hd(r)\cap \E\not=\emptyset$. 
This holds if and only if
$lit(\bd(r))\not\subseteq lit(\E)=\lit(\ua(\E))$ or $\ua(\hd(r))\cap
\ua(\E)\not=\emptyset$, which is equivalent to
$\nup(AIC(r))\not\subseteq\lit(\ua(\E))$ or $\hd(AIC(r))\cap \ua(\E)\not
=\emptyset$. This, however, is the definition of $AIC(r)$ closed under
$\ua(\E)$).  
~\hfill $\Box$

\begin{corollary}
\label{corollary:closeness}
Let $P$ be a proper revision program. 
A set $\E$ of revision literals is a minimal set closed under $P$ if and only if
$\ua(\E)$ is a minimal set closed under $AIC(r)$.
\end{corollary}
\textbf{Proof:} Straightforward from Lemma \ref{lemma:closeness}.~\hfill $\Box$

\begin{theorem}
\label{WRevisionWRepairs}
Let $P$ be a proper revision program. A set $\E$ of revision literals
is a (weak) revision (respectively, $P$-justified (weak) revision, $P$-founded (weak) revision) of $\I$ wrt $P$ if and only if $\ua(\E)$
is a (weak) repair (respectively, justified (weak) repair, founded (weak) repair) for $\<\I,AIC(P)\>$.
\end{theorem}
\textbf{Proof:} \\
(1) A set $\E$ of revision literals is a weak revision 
of $\I$ wrt $P$ if and only if $\ua(\E)$ is a weak repair for 
$\<\I,AIC(P)\>$.

\noindent
Indeed, by the definition, $\E$ is a weak revision of 
$\I$ with respect to $P$ if and only if
\begin{enumerate}
\item[(a)] $\I\cap\{a\st\rin(a)\in\E\}=\emptyset$, 
$\{a\st\rout(a)\in\E\}\subseteq\I$; and
\item[(b)] $\I\oplus\E\models P$.
\end{enumerate}
Similarly, $\ua(\E)$ is a weak repair for $\<\I,AIC(P)\>$ if and
only if 
\begin{enumerate}
\item[(a)] $\I\cap\{a\st +a \in\ua(\E)\}=\emptyset$, $\{a\st -a \in\ua(\E)
\}\subseteq\I$; and
\item[(b)] $\I\circ\ua(\E)\models AIC(P)$.
\end{enumerate}
By our earlier comments, for every database $\J$, $\J\models P$ if and 
only if $\J\models AIC(P)$. Since $\I\oplus\E=\I\circ\ua(\E)$, the
assertion follows.

\smallskip
\noindent
(2) Next, we prove that $\E$ is a revision of $\I$ wrt $P$ if and only 
if $\ua(\E)$ is a repair for $\<\I,AIC(P)\>$.

\noindent
By (1), $\E$ is a weak revision of $\I$ wrt $P$ if and only if $\ua(\E)$
is a weak repair for $\<\I,AIC(P)\>$. Moreover, we have that the mapping
$\E\mapsto \ua(\E)$ is a bijection between sets of revision literals and
sets of update actions such that $\I\oplus\E\models P$ if and only if
$\I\circ\ua(\E)\models AIC(P)$. Thus, a set $\E$ of revision literals is
such that for each $\E'\subseteq\E$ the fact $\I\oplus\E'\models P$ 
implies $\E'=\E$ (minimality of $\E$) if and only if $\ua(\E)$ is a set
of update actions such that for each $\F'\subseteq \ua(\E)$ the fact 
$\I\circ\F'\models AIC(P)$ implies $\F'=\ua(\E)$ (minimality of $\ua(\E)$.

\smallskip
\noindent
(3) We now prove that $\E$ is a $P$-justified weak revision of 
$\I$ if and only if $\ua(\E)$ is a justified weak repair for 
$\<\I,AIC(P)\>$.

\smallskip
\noindent
$(\Rightarrow)$
Since $\E$ is a $P$-justified weak revision of $\I$, there exists a 
$P$-justified weak update of $\I$, say $\U$, such that $\E=\U\setminus
I(\I,\I\oplus\U)$. By the definition, $\U$ is consistent and it is a 
minimal set containing $I(\I,\I\oplus\U)$ and closed under $P$. 
It follows that the action set $\ua(\U)$ is consistent and,
by Corollary \ref{corollary:closeness}, it is a minimal set
containing $\ua(I(\I,\I\oplus\U))$ and closed under $AIC(P)$.
We now observe that $\ua(I(\I,\I\oplus\U))=\nef(\I,\I\circ \ua(\U))$.
Thus, $\ua(\U)$ is a justified action set for $\<\I,AIC(P)\>$ and
$\ua(\U)\setminus\nef(\I,\I\circ \ua(\U))=
\ua(\E)$ is a justified weak repair for $\<\I,AIC(P)\>$.

\smallskip
\noindent
$(\Leftarrow)$
There exists a justified action set for $\<\I,AIC(P)\>$, say $\U$, such 
that $\ua(\E)=\U\setminus \nef(\I,\I\circ\U)$.
The action set $\U$ is consistent, contains $ne(\I,\I\circ\U)$ and 
it is closed under $AIC(P)$. By our comments above, there is a set of
revision literals $\V$ such that $\ua(\V)=\U$. Moreover, 
$\nef(\I,\I\circ\U)=\ua(I(\I,\I\oplus \V))$. It follows that the set 
$\V$ is consistent and, by Corollary \ref{corollary:closeness}, it is
a minimal set containing $I(\I,\I\oplus \V)$ and closed under $P$.
Thus, $\V$ is a $P$-justified weak update for $\I$ and
$\V\setminus I(\I,\I\oplus \V)= \E$ is a $P$-justified weak revision 
for $\I$.

\smallskip
\noindent
(4) By (3) and by the argument we used in (2) to show that the minimality
of $\E$ is equivalent to the minimality of $\ua(\E)$, $\E$ is a 
$P$-justified revision of $\I$ if and only if $\ua(\E)$ is 
a justified repair for $\<\I,AIC(P)\>$.

\smallskip
\noindent
(5) Finally, we prove that $\E$ is a $P$-founded (weak) revision of 
$\I$ if and only if $\ua(\E)$ is a founded (weak) repair for 
$\<\I,AIC(P)\>$.

\smallskip
\noindent
$(\Rightarrow)$
Let $\E$ be a $P$-founded (weak) revision of $\I$. By (1) and (2), 
$\ua(\E)$ is a (weak) repair for $\<\I,AIC(P)\>$. 
Therefore, we have to show that $\ua(\E)$ is founded wrt
$\<\I,AIC(P)\>$. Let us consider an arbitrary element of $\ua(\E)$.
It is of the form $\ua(\alpha)$, for some revision literal $\alpha
\in\E$. 

Since $\E$ is $P$-founded wrt $\I$, there exists $r\in P$
such that $\I\oplus\E\models\bd(r)$, and $\I\oplus\E\models \gamma^D$,
for every $\gamma\in\hd(r)$ different from $\alpha$. Let $\rho$ be the 
corresponding active integrity constraint in $\aic(P)$, that is, $\rho
=\aic(r)$. Since $r$ is proper, $\lit(\bd(r))=\nup(\rho)$. Thus,
$\I\circ\ua(\E)\models\nup(\rho)$. Moreover, since $\hd(\rho)=
\ua(\hd(r))$, for every $\delta\in\hd(\rho)$ other than $\ua(\alpha)$,
$\I\circ\ua(E)\models \delta^D$.

Thus, $\ua(\alpha)$ is founded wrt $\<\I,AIC(P)\>$ and 
$\ua(\E)$ and so, $\ua(\E)$ is founded with respect to $\<\I,AIC(P)\>$. 

\smallskip
\noindent
$(\Leftarrow)$ This implication can be proved by a similar argument.
We omit the details.  ~\hfill$\Box$

The results of this section show that proper revision programs can be
interpreted as sets of active integrity constraints so that the
corresponding semantics match. However, it is easy to see that the 
mapping $\aic(\cdot)$ is a 
one-to-one and onto mapping between the collection of proper revision programs
and the collections of sets of active integrity constraints.
Thus, also conversely, sets of active integrity constraints can
be interpreted as revision programs.
\begin{example}
Let $\eta$ be the following set of active integrity constraints:
\[
\begin{array}{lrll}
r_1= & a, b, \nt c & \supset & -a|+c \\
r_2= & \nt d & \supset & +d \\
r_3= & a & \supset & \bot \\
\end{array}
\]
The corresponding revision program is:
\[
\begin{array}{lrll}
 \rho_1= &\rout(a)|\rin(c) & \leftarrow & \rin(b) \\
 \rho_2= &\rin(d) & \leftarrow &  \\
 \rho_3= &\bot & \leftarrow &\rin(a).  \\
\end{array}
\]
\end{example}

The correspondence between sets of active integrity constraints and proper
revision programs allows us to adapt results from one setting to another 
and conversely. Moreover, in many cases, once we have a result for proper
revision programs, we can lift it to the general case, too. For instance, 
as in the case of sets of active integrity constraints and justified (weak)
repairs, a special structure of a revision program with respect to the 
original database ensures minimality of justified weak revisions. 
Specifically, we have the following corollary of Theorem
\ref{td}.

\begin{theorem}
\label{theorem::nrp}
Let $\I$ be a database and $P$ a revision program such that for each 
revision literal $\alpha$ appearing in the head of a rule in $P$, $\I \models 
\alpha^D$. If $\E$ is a $P$-justified weak revision for $\I$, then $\E$ is a 
$P$-justified revision for $\I$.
\end{theorem}
\textbf{Proof:} (Sketch) Clearly, the properized version $P'$ of $P$ also 
satisfies the assumption of the theorem. By the correspondence results 
between proper revision programs and sets of aic's, it follows that
$E$ is a $P'$-justified revision for $\I$. As $\I$ has the same 
justified revisions with respect to $P'$ and $P$, the result follows. 
\hfill$\Box$

Moreover, for normal revision programs justified weak revisions
are justified revisions no matter what the initial database, as stated in
the following corollary to Theorem \ref{tn}. The argument is essentially
the same as the one above and we omit it.

\begin{theorem}
\label{tn:rp}
Let $\I$ be a database and $P$ a normal revision program.
If $\E$ is a $P$-justified weak revision for $\I$, then $\E$ is a justified
revision for $\I$.
\end{theorem}

\section{Computation and Complexity Results for Revision Programming}
\label{summaryRev}
Thanks to the equivalence properties reported in Section \ref{equivalence} we can
derive the results about computation and complexity for revision programming
from the corresponding results for active integrity constraints 
presented in Section \ref{complexity}.

\begin{theorem}
Let $\I$ be a database and $P$ a  normal revision program.
Then checking if there exists a $P$-justified revision 
($P$-justified weak revision, respectively) for $\I$ is an NP-complete
problem.
\end{theorem}
\textbf{Proof.} By Theorem \ref{properization:thm} we know that
this problem is equivalent to check 
if there exists a $P'$-justified revision 
($P'$-justified weak revision, respectively) for $\I$ where $P'$ is the properized version of
$P$ that can be computed in polyno\-mial time. The result follows from Theorems \ref{jrc} and Theorem \ref{WRevisionWRepairs}.
~\hfill$\Box$

\begin{theorem}
Let $\I$ be a database and $P$ a  revision program.
Then checking if there exists a $P$-justified revision 
($P$-justified weak revision, respectively) for $\I$ is a $\Sigma_2^P$-complete problem.
\end{theorem}
\textbf{Proof.} 
By Theorem \ref{properization:thm} we know that
this problem is equivalent to check 
if there exists a $P'$-justified revision 
($P'$-justified weak revision, respectively) for $\I$ where $P'$ is the properized version of
$P$ that can be computed in polynomial time. 
The result follows from Theorems \ref{jwrsp2}, \ref{jrsp2} and \ref{WRevisionWRepairs}.
~\hfill$\Box$

\begin{theorem}
Let $\I$ be a database and $P$ a  revision program.
Then checking if there exists a $P$-founded revision ($P$-founded weak revision, respectively)
for $\I$ is a $\Sigma_2^P$-complete (NP-complete, respectively) problem.
\end{theorem}
\textbf{Proof.}
By Theorem \ref{properization:thm} we know that
this problem is equivalent to check 
if there exists a $P'$-founded revision ($P'$-founded weak revision, respectively) 
for $\I$ where $P'$ is the properized version of
$P$ that can be computed in polynomial time. 
The result follows from complexity results by Caroprese et al. \citeyear{CaroGZ}
and Theorem \ref{WRevisionWRepairs}.
~\hfill$\Box$

We summarize the complexity results obtained in this section in Figure
\ref{fig1b}.

\begin{figure}
\begin{center}
{\tiny
\begin{tabular}{ccccccccc}
& &  & & $\bFRev(\I,P^n)$ & & & &\\ 
& &  & & [$\Sigma_P^2$-c] & & & & \\
& &  & & \veq & & & & \\
$\bJRev(\I,P^n)$ & $\subseteq$ & $\bJRev(\I,P)$ & $\subseteq$ & $\bFRev(\I,P)$ & $\subseteq$ & $\bRev(\I,P)$ & $=$ & $\bRev(\I,P^n)$ \\
\mbox{[$NP$-c]} & & \mbox{[$\Sigma_P^2$-c]} & & \mbox{[$\Sigma_P^2$-c]} & & \mbox{[$NP$-c]} & & \mbox{[$NP$-c]}\\ 
\veq & & \vinc & & \vinc & & \vinc & & \vinc\\
$\bJWRev(\I,P^n)$&$\subseteq$ & $\bJWRev(\I,P)$ & $\subseteq$ & $\bFWRev(\I,P)$ & $\subseteq$ & $\bWRev(\I,P)$ & $=$ & $\bWRev(\I,P^n)$\\
\mbox{[$NP$-c]}&$ $ & \mbox{[$\Sigma_P^2$-c]} & $ $ & \mbox{[$NP$-c]} & $ $ & \mbox{[$NP$-c]} & $ $ & \mbox{[$NP$-c]}\\
& &  & & \veq & & & & \\
& &  & & $\bFWRev(\I,P^n)$ & & & & \\
& &  & & \mbox{[$NP$-c]} & & & & 
\end{tabular}
}
\caption{Complexity results for the semantics of revision programs} 
\label{fig1b}
\end{center}
\end{figure}

We note that comments we made at the end of Section \ref{summaryRP} apply 
here as well.
In a nutshell, a semantics of justified revisions, reflecting the principles 
of groundedness (no circular ``self-justifications'') and minimality of 
change, seems to be well motivated and so most appealing for applications.
However, as we pointed out earlier, it may be too restrictive. Thus, in all
these cases, when consistency of a database needs to be restored and justified
revisions do not exist, other semantics may provide an acceptable solution.
The discussion of that issue, involving also computational complexity 
trade-offs, follows essentially the same line as that in Section 
\ref{summaryRP}.

\section{Shifting Theorem}
\label{shifting}

In this section we study the \emph{shifting} transformation
\citep{mt94c}. The process consists of transforming an instance 
$\<\I,\eta\>$ of the database 
repair problem to a syntactically isomorphic instance $\<\I',\eta'\>$
by changing integrity constraints to reflect the ``shift'' of $\I$ into
$\I'$. A semantics for database repair problem has the \emph{shifting 
property} if the repairs of the ``shifted'' instance of the database 
update problem are precisely the results of modifying the repairs of 
the original instance according to the shift from $\I$ to $\I'$. The 
shifting property is important. If a semantics of database updates has 
it, the study of that semantics can be reduced to the case when the input
database is the empty set, a major conceptual simplification.

\begin{example}
\label{ex41}
Let $\I=\{a,b\}$ and let $\eta_5 = \{a, b \supset -a|-b\}$. There are two
founded repairs for $\<\I,\eta_5\>$: $\E_1=\{-a\}$ and $\E_2=\{-b
\}$. Let $\W=\{a\}$. We will now ``shift'' the instance $\<\I,\eta_5\>$
with respect to $\W$. To this end, we will first modify $\I$ by changing
the status in $\I$ of elements in $\W$, in our case, of $a$. Since
$a\in \I$, we will remove it. Thus, $\I$ ``shifted'' with respect to 
$\W$ becomes $\J=\{b\}$. Next, we will modify $\eta_5$ correspondingly,
replacing literals and update actions involving $a$ by their duals. That
results in $\eta'_5=\{\nt a, b\supset +a| -b\}$. One can check that the 
resulting instance $\<\J,\eta'_5\>$ of the update problem has two founded
repairs: $\{+a\}$ and $\{-b\}$. Moreover, they can be obtained from the
founded repairs for $\<\I,\eta_5\>$ by consistently replacing $-a$ with
$+a$ and $+a$ with $-a$ (the latter does not apply in this example).
In other words, the original update problem and its shifted version are
isomorphic.
~\hfill$\Box$
\end{example}

The situation presented in Example \ref{ex41} is not coincidental.
In this section we will show that the semantics of (weak) repairs, 
founded (weak) repairs and justified (weak) repairs satisfy the shifting
property. To facilitate the presentation, we placed proofs of all the
results in the appendix.

We start by observing that \emph{shifting} a database $\I$ to a 
database $\I'$ can be modeled by means of the symmetric difference 
operator. Namely, we have $\I'=\I\div\W$, where $\W=\I\div\I'$. This 
identity shows that one can shift any database $\I$ into any database 
$\I'$ by forming a symmetric difference of $\I$ with some set $\W$ of atoms  
(specifically, $\W=\I\div\I'$). We will now extend the operation of
shifting a database with respect to $\W$ to the case of literals, update
actions and integrity constraints. To this end, we introduce a
\emph{shifting} operator $T_\W$. 

\begin{definition}
Let $\W$ be a database and $\ell$ a literal or an update action. We 
define
\[
T_{\cal W}(\ell)=	\left\{\begin{array}{lll}
			\ell^D &\mbox{\ \ if} & \mbox{the atom of $\ell$ is 
                                                  in $\W$}\\
 			\ell   &\mbox{\ \ if} & \mbox{the atom of $\ell$ is 
                                                  not in $\W$}\\
 	 		\end{array}\right.
\]
and we extend this definition to sets of literals or
update actions, respectively. 

Furthermore, if $op$ is an operator on sets of literals or update actions
(such as conjunction or disjunction), for every set $X$ of literals or 
update actions, we define
\[
T_{\cal W}(op(X))=op(T_{\cal W}(X)).
\]

Finally, for an active integrity constraint $r= \phi\supset\psi$, we set
\[
T_{\cal W}(r)=T_{\cal W}(\phi) \supset T_{\cal W}(\psi).
\]
We extend the notation to sets active integrity constraints in the 
standard way. 
~\hfill $\Box$
\end{definition}

To illustrate the last two parts of the definition, we note that when $op$ 
stands for the conjunction of a set of literals and $X=\{L_1,\dots,L_n
\}$, where every $L_i$ is a literal, $T_{\cal W}(op(X))=op(T_{\cal
W}(X))$ specializes to
\[
T_{\cal W}(L_1,\dots,L_n)=T_{\cal W}(L_1),\dots,T_{\cal W}(L_n). 
\]
Similarly, for an active integrity constraint
\[
r= L_1,\dots,L_n\supset\alpha_1|\ldots|\alpha_m
\]
we obtain
\[
T_\W(r) = T_{\cal W}(L_1),\dots,T_{\cal W}(L_n)\supset T_{\cal
W}(\alpha_1)|\dots|T_{\cal W}(\alpha_m).
\]
To summarize, we overload the notation $T_\W$ and interpret it based on the 
type of the argument.

\begin{theorem}[\textsc{Shifting theorem for (weak) repairs and founded
repairs}]
\label{theorem:shifting}
Let $\I$ and $\W$ be databases. For every set $\eta$ of active integrity 
constraints and for every consistent set $\E$ of update actions, we have
\begin{enumerate}
\item $\E$ is a weak repair for $\langle\I,\eta\rangle$ if and only if 
$T_{\W}(\E)$ is a weak repair for $\langle \I\div\W,T_{\W}(\eta)\rangle$
\item $\E$ is a repair for $\langle\I,\eta\rangle$ if and only if 
$T_{\W}(\E)$ is a repair for $\langle \I\div\W,T_{\W}(\eta)\rangle$
\item $\E$ is founded for  $\langle\I,\eta\rangle$ if and only if 
$T_{\W}(\E)$ is founded for $\langle \I\div\W,T_{\W}(\eta)\rangle$
\item $\E$ is a founded (weak) repair for $\langle\I,\eta\rangle$ if 
and only if $T_{\W}(\E)$ is a founded (weak) repair for $\langle \I
\div\W,T_{\W}(\eta)\rangle$
\item $\E$ is an
justified (weak) repair for $\langle\I,\eta\rangle$ if and only
if $T_{\W}(\E)$ is a justified (weak) repair for $\<\I\div\W,T_{\W}(\eta)\>$.
\end{enumerate}
\end{theorem}

Theorem \ref{theorem:shifting} implies that in the context of (weak) 
repairs, founded (weak) repairs or justified (weak) repairs, an instance
$\<\I,\eta\>$ of the database update problem can be shifted to the 
instance with the empty initial database. That property can simplify
studies of these semantics as well as the development of algorithms
for computing repairs and for consistent query answering, as it allows
us to eliminate one of the parameters (the initial database) from 
considerations. In many cases it also allows us to relate semantics 
of database repairs to some semantics of logic programs with negation.
Formally, we have the following corollary.

\begin{corollary}
\label{fc}
Let $\I$ be a database and $\eta$ a set of active integrity constraints.
Then $\E$ is a weak repair (repair, founded weak repair, founded repair,
justified weak repair, justified repair, respectively) for $\<\I,\eta\>$
if and only if $T_\I(\E)$ is a weak repair (repair, founded weak repair,
founded repair, justified weak repair, justified repair, respectively)
for $\<\emptyset,T_{\I}(\eta)\>$.
\end{corollary}

The concept of of shifting can also be stated for revision programming.
First, we note that the operator $T_{\W}(\cdot)$ defined above can be 
extended to revision literals, revision rules and revision programs. 
Its formal definition and many properties have been presented by Marek 
et al. \citeyear{mpt98}. The following theorem gathers those results,
as well as their extensions to the case of new semantics we introduced
in our paper. 

\begin{theorem}\textsc{(Shifting theorem for revision programs)}
\label{theorem:shiftingRevisions}
Let $\I$ and $\W$ be databases. For every revision program $G$ 
and every consistent set $\E$ of revision literals, we have
\begin{enumerate}
\item $\E$ is a (weak) revision for $\I$ with respect to $G$ if and 
only if $T_{\W}(\E)$ is a (weak) revision for $\I$ with respect to 
$T_{\W}(G)$
\item $\E$ is a $G$-justified (weak) revision for $\I$ if and only 
if $T_{\W}(\E)$ is a $T_{\W}(G)$-justified (weak) revision for $\I$
\item $\E$ is a $G$-founded (weak) revision for $\I$ if and only if 
$T_{\W}(\E)$ is a $T_{\W}(G)$-founded (weak) revision for $\I$
\end{enumerate}
\end{theorem}

\section{Conclusion}

In the paper we studied two formalisms for describing policies on
enforcing integrity constraints on databases in the presence of 
preferences on alternative ways to do so: \emph{active integrity 
constraints} \citep{CaroGZ} and \emph{revision programming} \citep{mt94c}. 

The original semantics proposed for active integrity constraints is based
on the concept of a \emph{founded repair}. A founded repair is a set of 
\emph{update actions} (\emph{insertions} and \emph{deletions}) to be 
performed over the database in order to make it consistent, that is 
minimal and \emph{supported} by active integrity constraints. The 
original semantics for revision programs is based on the concept of 
\emph{justified revision}. A justified revision is a set of \emph{revision 
literals} that can be inferred by means of the revision program and by 
the \emph{inertia set}, that is the set of all atoms that do not
change their state of \emph{presence} in or \emph{absence} from a 
database during the revision process.

We proved that in the context of their original semantics, these two
formalisms differ. That is, under some natural interpretation of revision
programs as sets of active integrity constraints the set of repairs 
corresponding to justified revisions is contained in the set of founded
repairs (and the containment is, in general, proper). This observation 
demonstrated that basic intuitions behind the two semantics are essentially
different and opened a possibility of expanding each formalism by semantics
grounded in the ideas developed in the other one.

Following this direction, we introduced a new semantics for active 
integrity constraints, based on ideas underlying revision programming
and, conversely, a new semantics for revision programs based on 
intuitions behind founded repairs.  With the new semantics available, 
we showed that the interpretation of revision programs as sets of active
integrity constraints, mentioned above, establishes a precise match 
between these two formalism: it preserves their semantics once they are 
correctly aligned. In other words, we proved that the two formalisms are 
equivalent through a simple modular (rule-wise) syntactic transformation.
That offers a strong indication of the adequacy of each formalism as the
foundation for declarative specifications of policies for enforcing
integrity constraints. Moreover, the broad frameworks of semantics we 
have available in each case provide us with means of handling the problem 
of ``non-executability'' of the policies encoded into integrity constraints
under a particular semantics: once that turns out to be the case, one can
chose to select a less restrictive one.

For each formalism and each semantics we established the complexity of 
the basic existence of repair (revision) problem. Furthermore, we proved
that each formalism and each semantics satisfies the \emph{shifting 
property}. Shifting consists of transforming an instance of a database 
repair problem to another syntactically isomorphic instance by changing 
active integrity constraints or revision programs to reflect the 
``shift'' from the original database to the new one. 

These latter results are essential for relating repair (revision)
formalisms we studied with logic programming and, specifically, with
programs that generalize standard disjunctive logic programs by
allowing default literals also in the heads of disjunctive rules
(the Lifschitz-Woo programs \citep{lw92}; cf. work by Marek at al.
\citeyear{mpt98} and Pivkina \citeyear{piv01} for some early results
exploiting shifting to relate revision and logic programming).

Our work opens and forms the foundation for several research directions. 
The first of them concerns implementations of algorithms for computing 
repairs (revisions) under the semantics discussed here in the first-order
setting covering built-in predicates and aggregates. 
An important aspect of that research is to identify classes of
databases and integrity constraints, for which the existence and the uniqueness
of repairs (revisions) of particular types is assured.

The second problem concerns consistent query answering 
in the setting of active integrity constraints. The problem is to compute 
answers to queries to a database that is inconsistent with respect to its 
active integrity constraints without computing the repairs explicitly, 
thus extending the approach of consistent query answering 
\citep{ArenasBC99,ArenasBC03,Chomicki07} to the setting of active 
integrity constraints.

Next, there is the question whether a still narrower 
classes of repairs could be identified based on the analysis of all active 
integrity constraints (revision program rules) that would resolve conflicts 
among them (multiple possible repairs or revisions result precisely from
the need to choose which constraint or rule to use when several are
applicable) either based on their specificity (an approach used with some 
success in default logic) or on explicit rankings of the relative importance 
of active integrity constraints and revision rules.

Finally, we note that all the semantics discussed in the paper give rise
to knowledge base operators that could be analyzed from the standpoint of
Katsuno-Mendel\-zon postulates. To this end, we observe that we can view a 
set of databases as the set of models of some formula and so, as a knowledge
base in the sense of Katsuno and Mendelzon \citep{DBLP:conf/kr/KatsunoM91}.
Let $\eta$ be a set of active integrity constraints and, for the sake of 
illustration, let us focus our attention on the semantics of justified
repairs. Given a set of databases (a knowledge base), $\I$, we can assign 
to it another set of databases (knowledge base), $\I'$, consisting of 
all $\eta$-justified repairs of all databases in $\I$. In that way we 
obtain a knowledge base update operator determined by $\eta$ and the 
semantics justified repairs. It is an interesting problem to determine 
which of the Katsuno-Mendelzon postulates are satisfied by that operator 
(and by the other ones that arise by choosing a different update 
semantics).

\section*{Acknowledgments}
The authors thank anonymous reviewers for many insightful comments that
resulted in substantial improvements to the original manuscript.
This work was partially supported by the NSF grants IIS-0325063 and 
IIS-0913459, and the KSEF grant KSEF-1036-RDE-008.

\bibliography{Journal}

\section*{Appendix}

We present here the proofs of the two shifting theorems. The proofs 
are based on several auxiliary results.

\begin{lemma}
\label{fact:R}
Let $\W$ be a database.
\begin{enumerate}
\item\label{R.1} 
For every update action $\alpha$, $T_\W(\lit(\alpha))=\lit(T_\W(
 \alpha))$
\item\label{R.2}
For every set $A$ of literals (update actions, active integrity 
constraints, respectively) $T_{\cal W}(T_{\cal W}(A))=A$
\item\label{R.3}
For every consistent set $\A$ of literals (update actions, 
respectively), $T_\W(\A)$ is consistent
\item\label{R.4} 
For every databases $\I$ and $\R$, $T_\W(\nef(\I,\R))= \nef(\I\div\W,
\R\div\W)$
\item\label{R.5}
For every active integrity constraint $r$, $\nup(T_\W(r))=T_\W(\nup(r
))$.
\end{enumerate}
\end{lemma}
\textbf{Proof:} 
(\ref{R.1}) - (\ref{R.3}) follow directly from the 
definitions. We omit the details.

\smallskip
\noindent
\begin{enumerate}
\setcounter{enumi}{3}
\item
Let $\alpha\in\nef(\I\div\W,\R\div\W)$. If $\alpha=+a$, then it
follows that $a\in(\I\div\W)\cap(\R\div\W)$. Let us assume that
$a\in\W$. Then $a\notin\I\cup\R$ and, consequently, $-a\in\nef(\I,\R)$.
Since $a\in\W$, $+a=T_\W(-a)$. Thus, $\alpha\in T_\W(\nef(\I,\R))$. 
The case when $\alpha=-a$ can be dealt with in a similar way. It follows
that $\nef(\I\div\W,\R\div\W)\subseteq T_\W(\nef(\I,\R))$.

Let $\I'=\I\div\W$ and $\R'=\R\div\W$. Then $\I=\I'\div\W$,
$\R=\R'\div\W$ and, by applying the inclusion we just proved to 
$\I'$ and $\R'$, we obtain
\[
\nef(\I,\R)=\nef(\I'\div\W,\R'\div\W)\subseteq T_\W(\nef(\I',\R')).
\]
Consequently, 
\[
T_\W(\nef(\I,\R))\subseteq T_\W(T_\W(\nef(\I',\R')))=\nef(\I\div\W,\R
\div\W).
\]
Thus, the claim follows.

\item
Let $L\in\nup(T_\W(r))$. We have $L\in\bd(T_\W(r))$
and $L^D\notin\lit(\hd(T_\W(r))$. Clearly, $\hd(T_\W(r))=T_\W(\hd(r))$
and $\bd(T_\W(r))=T_\W(\bd(r))$. Thus, $L\in
T_\W(\bd(r))$ and $L^D\notin T_\W(\hd(r))$. Consequently, $T_\W(L)\in
\bd(r)$. Moreover, since $T_\W(L^D)=(T_\W(L))^D$, $(T_\W(L))^D\notin
\hd(r)$. It follows that $T_\W(L)\in\nup(r)$ and so, $L\in
T_\W(\nup(r))$. Hence, $\nup(T_\W(r))\subseteq T_\W(\nup(r))$.

Applying this inclusion to an active integrity constraint $s=T_\W(r)$,
we obtain $\nup(r)\subseteq T_\W(\nup(T_\W(r)))$. This, in turn,
implies
$T_\W(\nup(r))\subseteq T_\W(T_\W(\nup(T_\W(r))))=\nup(T_\W(r))$.
Thus, the equality $\nup(T_\W(r))= T_\W(\nup(r))$ follows.
~\hfill $\Box$
\end{enumerate}

\begin{lemma}
\label{fact:R-1}
Let $\I$ and $\W$ be databases and let $L$ be a literal or an update
action. Then $\I\models L$ if and only if $\I\div\W\models T_\W(L)$. 
\end{lemma}
\textbf{Proof:} ($\Rightarrow$) Let us assume that $\I\models L$.
If $L = a$, where $a$ is an atom, then $a\in \I$. There are two 
cases: $a\in\W$ and $a\notin\W$. In the first case, $a\notin \I\div\W$ 
and $T_\W(a)= \nt a$. In the second case, $a\in\I\div\W$ and 
$T_\W(a)=a$. In each case, $\I\div\W\models T_\W(a)$, that is, $\I\div\W
\models T_\W(L)$. 

The case $L=\nt a$, where $a$ is an atom, is similar. First, we have that
$a\notin\I$. If $a\in\W$ then $a\in\I\div\W$ and $T_\W(\nt a)= a$. If
$a\not\in\W$ then $a\notin\I\div\W$ and $T_\W(\nt a)= \nt a$. In each 
case, $\I\div\W\models T_\W(\nt a)$, that is, $\I\div\W \models T_\W
(L)$.

\smallskip
\noindent
($\Leftarrow$) Let us assume that $\I\div\W\models T_\W(L)$. Then,
$(I\div\W)\div\W=\I$ and $T_\W(T_\W(L)) =L$. Thus, $\I\models L$
follows by the implication ($\Rightarrow$).
~\hfill$\Box$

\begin{lemma}
\label{fact:R1}
Let $\I$ and $\W$ be databases, and let $\U$ be a consistent set of 
update actions. Then $(\I\circ \U)\div\W = (\I\div\W)\circ T_\W(\U)$.
\end{lemma}
\textbf{Proof:} We note that since $\U$ is consistent, $T_\W(\U)$ is
consistent, too. Thus, \emph{both} sides of the identity are well
defined.

Let $a\in(\I\circ \U)\div\W$. If $+a\in T_\W(\U)$, then $a\in (\I\div\W)
\circ T_\W(\U)$. Thus, let us assume that $+a\notin T_\W(\U)$. We have two 
cases.\\
Case 1: $a\notin\W$. From the definition of $T_\W$, $+a\notin \U$. Since 
$a\in(\I\circ \U) \div\W$, $a\in\I\circ \U$ and, consequently, $a\in\I$
and $-a\notin \U$. Thus, $a\in(\I\div\W)$ and $-a\notin T_\W(\U)$
(otherwise, as $T_\W(-a)=-a$, we would have $-a\in\U$). 
Consequently, $a\in (\I\div\W)\circ T_\W(\U)$.\\
Case 2: $a\in\W$. From the definition of $T_\W$, $-a\notin \U$. Since
$a\in(\I\circ \U) \div\W$, $a\notin\I\circ\U$. Thus, $a\notin\I$ and 
$+a\notin \U$. It follows that $a\in\I \div\W$ and $-a\notin T_\W(\U)$
(otherwise we would have $+a\in\U$, as $T_\W(-a)=+a$, in this case).
Hence, $a\in(\I\div\W)\circ T_\W(\U)$.

If $a\notin (\I\circ \U)\div\W$, we reason similarly. If $-a\in T_\W(\U)$,
then $a\notin (\I\div\W) \circ T_\W(\U)$. Therefore, let us assume that
$-a\notin T_\W(\U)$. As before, there are two cases.\\
Case 1: $a\notin\W$ and thus $-a\notin \U$. Since $a\notin(\I\circ\U)
\div\W$, $a\notin\I\circ \U$ and, consequently, $a\notin\I$ and $+a\notin
\U$. Thus, $a\notin(\I\div\W)$ and $+a\notin T_\W(\U)$. Consequently, $a
\notin (\I\div\W)\circ T_\W(\U)$.\\
Case 2: $a\in\W$ and thus $+a\notin \U$. In this case, $a\in \I\circ 
\U$. Thus, $a\in\I$ and $-a\notin \U$. It follows that $a\notin\I\div\W$
and $+a\notin T_\W(\U)$. Hence, $a\notin(\I
\div\W)\circ T_\W(\U)$. ~\hfill$\Box$

\begin{lemma}
\label{lem:R0}
Let $\I$ and $\W$ be databases, $\U$ a consistent set of update actions,
and $L$ a literal or an action update. Then $\I\circ \U\models L$ if 
and only if $(\I\div\W) \circ T_\W(\U)\models T_\W(L)$.~\hfill$\Box$
\end{lemma}
\textbf{Proof}: 
By Lemma \ref{fact:R-1}, $\I\circ \U\models L$ 
if and only if $(\I\circ \U)\div\W\models T_\W(L)$. By Lemma
\ref{fact:R1}, the latter condition is equivalent to the condition 
$(\I\div\W)\circ T_\W(\U) \models T_\W(L)$. 
~\hfill$\Box$

\begin{lemma}
\label{proposition:shiftingJR}
Let $\I$ and $\W$ be databases. For every set $\eta$ of active integrity
constraints and for every set $\U$ of update actions, $\U$ is a justified
action set for $\<\I,\eta\>$ if and only if $T_{\W}(\U)$ is a
justified action set for $\<\I\div \W,T_{\W}(\eta)\>$.
\end{lemma}

\smallskip
\noindent
\textbf{Proof:}
($\Rightarrow$)
We have to prove that $T_{\W}(\U)$ is consistent, and minimal among
all supersets of $\nef(\I\div\W,(\I\div\W)\circ T_{\W}(\U))$ that are
closed under $T_{\W}(\eta)$.

Since $\U$ is a justified action set for $\<\I,\eta\>$, $\U$ is
consistent and $\nef(\I,\I\circ\U)\subseteq\U$. The former implies
that $T_{\W}(\U)$ is consistent (cf. Lemma \ref{fact:R}(\ref{R.1}
)). The latter implies that $\nef(\I\div\W,(\I\div\W)\circ T_{\W}(\U))
\subseteq T_\W(\U)$ (cf. Lemma \ref{fact:R}(\ref{R.2}) and
\ref{fact:R1}).

Next, we prove that $T_{\W}(\U)$ is closed under $T_{\W}(\eta)$. Let
$r$ be an active integrity constraint in $T_{\W}(\eta)$ such that
$\bd(r)$ is consistent, $\nup(r)\subseteq \lit(T_{\W}(\U))$. Then,
there exists $s\in\eta$ such that $r=T_{\W}(s)$.
By Lemma \ref{fact:R}(\ref{R.5}), $\nup(r)=T_{\W}(\nup(s))$.
As $T_{\W}(\nup(s)) \subseteq \lit(T_{\W}(\U))$, we have that $\nup(s)
\subseteq \lit(\U)$. Since $\U$ is closed under $s$, there exists
$\alpha\in\hd(s)$ such that $\alpha\in\U$. Thus, we obtain that $T_{\W
}(\alpha)\in T_{\W}(\hd(s))=\hd(r)$, and that $T_{\W}(\alpha)\in T_{\W
}(\U)$. Consequently, $\hd(r)\cap T_{\W}(\U)\not=\emptyset$. It follows
that $T_{\W}(\U)$ is closed under $r$ and so, also under $T_{\W}(\eta)$.

Finally, let us consider a set $\V$ of update actions such that
$\nef(\I\div\W,(\I\div\W)\circ T_{\W}(\U))\subseteq\V\subseteq T_\W(\U)$
and closed under $T_{\W}(\eta)$. By Lemma \ref{fact:R}(\ref{R.2})
and \ref{fact:R1}, $\nef(\I\div\W,(\I\div\W)\circ T_{\W}(\U))=T_{\W}(
\nef(\I,\I\circ\U))$. Thus, $\nef(\I,\I\circ\U)\subseteq T_{\W}(\V)
\subseteq \U$. From the fact that $\V$ is closed under $T_{\W}
(\eta)$ it follows that $T_{\W}(\V)$ is closed under $\eta$ (one can
show it reasoning similarly as in the previous paragraph). As $\U$ is
minimal in the class of supersets of $\nef(\I,\I\circ\U)$ closed under
$\eta$, $T_{\W}(\V)=\U$ and so, $\V=T_{\W}(\U)$. This completes the
proof of the implication ($\Rightarrow)$.

\smallskip
\noindent
($\Leftarrow$)
If $T_{\W}(\U)$ is a justified action set for $\<\I\div\W,T_\W(\eta)\>$,
the implication $(\Rightarrow)$ yields that $T_{\W}(T_{\W}(\U))=\U$ is
a justified action set for $\<(\I\div \W)\div\W=\I,\eta\>$.
~\hfill $\Box$

\medskip
\noindent
\textbf{Proof of Theorem \ref{theorem:shifting}}:
\begin{enumerate}
\item
Let us assume that $\E$ is a weak repair for 
$\langle\I,\eta\rangle$. It follows that $\E$ is consistent. Since
$\I\circ\E \models\eta$, by Lemma \ref{lem:R0}, $(\I\div\W)
\circ T_\W(\E) \models T_\W(\eta)$. The converse implication follows 
from the one we just proved by Lemma \ref{fact:R}(\ref{R.2}).

\item As before, it suffices to show only one implication. Let $\E$ be 
a repair for $\langle\I,\eta\rangle$. Then, $\E$ is a weak repair 
for $\langle\I,\eta\rangle$. By (1), $\E$ is a weak repair for $\langle
\I\div\W,T_{\W}(\eta)\rangle$. Let $\E'\subseteq T_\W(\E)$ be such that
$(\I\div\W)\circ \E'\models T_\W(\eta)$. It follows that $T_\W(\E')
\subseteq T_\W(T_\W(\E))=\E$. Since $\E$ is consistent, $T_\W(\E')$
is consistent, too. By Lemma \ref{lem:R0} and Lemma 
\ref{fact:R}(\ref{R.2}), since $(\I\div\W)\circ \E'\models T_\W(\eta)$, then
$\I\circ T_\W(\E') \models\eta$. Since $\E$ is a repair and $T_\W(\E')
\subseteq\E$, $T_\W(\E')=\E$. Thus, $\E'=T_\W(\E)$ and so, $T_\W(\E)$
is a repair for $\langle \I\div\W,T_{\W}(\eta) \rangle$. 

\item As in two previous cases, we show only one implication. Thus, let
us assume that $\E$ is founded for $\langle\I,\eta\rangle$. Let $\alpha
\in T_{\W}(\E)$. It follows that there is $\beta\in \E$ such that 
$\alpha=T_\W(\beta)$. Since $\E$ is founded with respect to $\<\I,
\eta\>$, there is an active integrity constraint $r$ such that
$\beta\in\hd(r)$,
$\I\circ\E\models\nup(r)$, and for every $\gamma\in\hd(r)\setminus
\{\beta\}$, $\I\circ\E\models\gamma^D$.

Clearly, the active integrity constraint $T_\W(r)$ belongs to $T_\W(
\eta)$ and $\alpha=T_\W(\beta)$ is an element of $\hd(T_\W(r))$.
By Lemma \ref{fact:R}(\ref{R.5}), we have $\nup(r)=\nup(T_\W(r))$.
Thus, by Lemma \ref{lem:R0}, $(\I\div\W)\circ T_\W(\E)\models
\nup(T_\W(r))$. Next, let $\gamma\in\hd(T_\W(r))\setminus\{\alpha\}$.
Then, there is $\delta\in \hd(r)\setminus\{\beta\}$ such that 
$\gamma=T_\W(\delta)$. Since $\I\circ\E\models\gamma^D$, it follows
that $(\I\div\W)\circ T_\W(\E)\models T_\W(\delta^D)$, that is,
$(\I\div\W)\circ T_\W(\E)\models\gamma^D$. Thus, $\alpha$ is founded 
with respect to $\<\I\div\W, T_\W(\eta)\>$ and $T_\W(\E)$ and 
$T_\W(\E)$ is founded with respect to $\<\I\div\W, T_\W(\eta)\>$.

\item This property is a direct consequence of (1), (2), and (3).
\item If $\E$ is a justified weak repair for
$\<\I,\eta\>$, then $\E\cap\nef(\I,\I\circ\E) =\emptyset$ and $\E\cup\nef(\I,
\I\circ\E)$ is a justified action set for $\<\I,\eta\>$ (Theorem
\ref{cor-930}). It follows that $T_\W(\E)\cap T_\W(\nef(\I,\I\circ\E))=
\emptyset$. Moreover, by Lemma \ref{proposition:shiftingJR}, $T_\W(\E\cup
\nef(\I,\I\circ\E))$ is a justified action set for $\<\I\div
\W,T_\W(\eta)\>$.

We have $T_\W(\nef(\I,\I\circ\E))=\nef(\I\div\W,(\I\div\W)\circ T_{\W}
(\E))$. Thus, again by Theorem \ref{cor-930}, $T_\W(\E)$ is a
justified weak repair for $\<\I\div\W,T_\W(\eta)\>$.

If $\E$ is a justified repair for $\<\I,\eta\>$, then our argument shows
that $T_\W(\E)$ is a justified weak repair for $\<\I\div\W,T_\W(\eta)\>$.
Moreover, since $\E$ is a repair for $\I$, by Theorem
\ref{theorem:shifting}(2) we have that $T_\W(\E)$ is a repair for $\I
\div\W$. It follows that $T_\W(\E)$ is a justified repair
for $\<\I\div\W,T_\W(\eta)\>$. The other implication can now be argued
in the same way as in several other similar cases in the paper.
~\hfill $\Box$
\end{enumerate}

\medskip
\noindent
\textbf{Proof of Corollary \ref{fc}:}
The assertion follows directly from Theorem \ref{theorem:shifting}.
~\hfill $\Box$ 

Next we turn to the shifting properties of revision programs. We will
derive Theorem \ref{theorem:shiftingRevisions} from Theorem 
\ref{theorem:shifting}. To this end we need one more lemma. 

\begin{lemma}
\label{lemma:TProperties}
Let $\I$ and $\W$ be databases, $\E$ a set of revision literals,
$G$ a revision program and $P$ a proper revision program.
Then  
$T_{\W}(\prop(G))=\prop(T_{\W}(G))$,
$T_{\W}(\ua(\E))=\ua(T_{\W}(\E))$ and 
$T_{\W}(\aic(P))=aic(T_{\W}(P))$.
\end{lemma}		
\textbf{Proof:}
Straightforward from
the definitions of $\prop(\cdot)$, $T_{\W}(\cdot)$, $\ua(\cdot)$ and 
$\aic(\cdot)$. ~\hfill $\Box$

\medskip
\noindent
\textbf{Proof of Theorem \ref{theorem:shiftingRevisions}:}
Let $P=\prop(G)$ (that is the ``properized'' version of $G$). The
following properties are equivalent:
\begin{enumerate}
\item $\E$ is a (weak) revision for $\I$ 
with respect to $G$ (respectively, $G$-justified (weak) revision for
$\I$, $G$-founded (weak) revision for $\I$)
\item $\E$ is a (weak) revision for $\I$ with respect to $P$
(respectively, $P$-justified (weak) revision for $\I$, $P$-founded 
(weak) revision for $\I$)
\item $\ua(\E)$ is a (weak) repair (respectively, justified (weak) 
repair, founded (weak) repair) for $\<\I,AIC(P)\>$
\item $T_{\W}(\ua(\E))$ is a (weak) repair (respectively, justified
(weak) repair, founded (weak) repair) for $\langle \I\div\W,T_{\W}
(AIC(P))\rangle$
\item $T_{\W}(\E)$ is a (weak) revision for $\I\div\W$ with respect to
$T_{\W}(P)$ (respectively, $T_{\W}(P)$-justified (weak) revision for 
$\I\div\W$, $T_{\W}(P)$-founded (weak) revision for $\I\div\W$)
\item $T_{\W}(\E)$ is a (weak) revision for $\I\div\W$ with respect to
$T_{\W}(G)$ (respectively, $T_{\W}(G)$-justified (weak) revision for 
$\I\div\W$, $T_{\W}(G)$-founded (weak) revision for $\I\div\W$).
\end{enumerate}
Indeed, (1) and (2) are equivalent by Theorem \ref{properization:thm},
(2) and (3) are equivalent by Theorem \ref{WRevisionWRepairs}, (3) and (4) --- by Theorems
6 and 7 of [8] 
(\emph{the shifting theorem for (weak) repairs, founded (weak) repairs and justified (weak) repairs}). 
Next, (4) and (5)
are equivalent by Theorem \ref{WRevisionWRepairs}, as well as Lemma
\ref{lemma:TProperties}, and (5) and (6) --- by 
Theorem \ref{properization:thm} and Lemma
\ref{lemma:TProperties}.
Thus, the assertion follows. 
 ~\hfill$\Box$

\end{document}